\documentclass[galaxies,review,accept,pdftex,moreauthors]{Definitions/mdpi} 
\firstpage{1} 
\pubvolume{1}
\issuenum{1}
\articlenumber{0}
\pubyear{2025}
\copyrightyear{2024}
\externaleditor{Alberto Sadun} 
\datereceived{20 September 2023} 
\daterevised{7 November 2024} 
\dateaccepted{16 December 2024} 
\datepublished{ } 
\hreflink{https://doi.org/} 

\usepackage{longtable}
\newcommand{\ao}{Appl. Opt.}
\newcommand{\nat}{Nature}
\newcommand{\apj}{Astrophys. J.}
\newcommand{\apjl}{Astrophys. J. Lett.}
\newcommand{\apjs}{Astrophys. J. Suppl.}
\newcommand{\araa}{Annu. Rev. Astron. Astrophys.} 
\newcommand{\aap}{Astron. Astrophys.}

\newcommand{\mnras}{Mon. Not. R. Astron. Soc.}

\newcommand{\aj}{Astron. J.}
\newcommand{\pasa}{Pub. Astron. Soc. Aust.}
\newcommand{\pasp}{Pub. Astron. Soc. Pac.}
\newcommand{\pasj}{Pub. Astron. Soc. Japan}

\newcommand{\aapr}{Astron. Astrophys. Rev.}

\newcommand{\teff}{\text{T}_{\text{eff}}}
\newcommand{\teffpl}{\text{T}_{\text{eff,p}}}

\Title{High-Contrast Imaging: Hide and Seek with Exoplanets}

\TitleCitation{High-Contrast Imaging: Hide and Seek with Exoplanets}

\Author{Riccardo Claudi$^{1, 2,}$*$^{,\dagger}$\orcidA{} and Dino Mesa $^{3,\dagger}$}

\AuthorNames{Riccardo Claudi, Dino Mesa}

\AuthorCitation{Claudi, R.; Mesa, D.}

\address{%
$^{1}$ \quad INAF---Astronomical Observatory of Padova, Vicolo Osservatorio, 5, 35122 Padova, Italy; riccardo.claudi@inaf.it \\
$^{2}$ \quad Dipartimento di Matematica e Fisica, Universit\`a Roma Tre, Via della Vasca Navale 84, 00146 Roma, Italy \\
 $^{3}$ \quad INAF---Astronomical Observatory of Padova, Vicolo Osservatorio, 5, 35122 Padova, Italy; dino.mesa@inaf.it
 }

\corres{Correspondence: riccardo.claudi@inaf.it}

\firstnote{These authors contributed equally to this work.}

\abstract{So far, most of the about 5700 exoplanets have been discovered mainly with radial velocity and transit methods. These techniques are sensitive to planets in close orbits, not being able to probe large star--planet separations. $\mu$-lensing is the indirect method that allows us to probe the planetary systems at the snow-line and beyond, but it is not a repeatable observation. On the contrary, direct imaging (DI) allows for the detection and characterization of low mass companions at wide separation (\mbox{$\leq$ 5--6 au}). The main challenge of DI is that a typical planet--star contrast ranges from $10^{-6}$, for a young Jupiter in emitted light, to $10^{-9}$ for Earth in reflected light.
In the last two decades, a lot of efforts have been dedicated to combining large (D $\geq$ 5 m) telescopes (to reduce the impact of diffraction) with coronagraphs and high-order adaptive optics (to correct phase errors induced by atmospheric turbulence), with sophisticated image post-processing, to reach such a contrast between the star and the planet in order to detect and characterize cooler and closer companions to nearby stars. Building on the first pioneering instrumentation, the second generation of high-contrast imagers, SPHERE, GPI, and SCExAO, allowed us to probe hundreds of stars (e.g., 500--600 stars using SHINE and GPIES), contributing to a better understanding of the demography and the occurrence of planetary systems. The DI offers a possible clear vision for studying the formation and physical properties of gas giant planets and brown dwarfs, and the future DI (space and ground-based) instruments with deeper detection limits will enhance this vision.
In this paper, we briefly review the methods, the instruments, the main sample of targeted  stars, the remarkable results, and the perspective of this rising technique. }


\begin{document}


\section{Introduction}
\label{sec:intro}
Perhaps Dyson and collaborators, having observed several bright stars of the Hyades during the Sun's eclipse in 1919, went on to perform the first high-contrast astronomical observation. The total Solar eclipse allowed us to observe the Hyades stars over the corona light background within a few arcsecs of the Sun and lets the observers measure the variation of the apparent position of these stars, confirming the general relativistic prediction \citep{dysonetal1920rspta220_291}. Also, if Dyson and collaborators exploited a total solar eclipse, they would be able to observe stars that were at least $10^{12}$ fainter than the Sun.
However, we had to wait until 2004 to obtain the first direct observation of a planetary-mass companion (PMC), which was taken with VLT/NACO by \citet{chauvinetal2004aa425_l29}, while in 2008, \citet{maroisetal2008sci322_1348} announced the first exoplanetary system (three planets) found orbiting the young A5V star HR8799, using direct imaging. This detection was fully 13 years after the first exoplanet orbiting a Solar-like star (95 Peg b\cite{mayorandqueloz1995nature378_355}) was discovered, and~after that, more than 400 new worlds were indirectly discovered by radial velocity, transit, and~microlensing techniques. This delay occurred mainly due to the extraordinary efforts that are necessary for the direct imaging techniques to overcome the difficulties imposed {by astrophysics (planet--star contrast), physics (diffraction and scattering), and~engineering (speckle noise reduction)}.

\begin{figure}[H]
\includegraphics[width=10.5 cm]{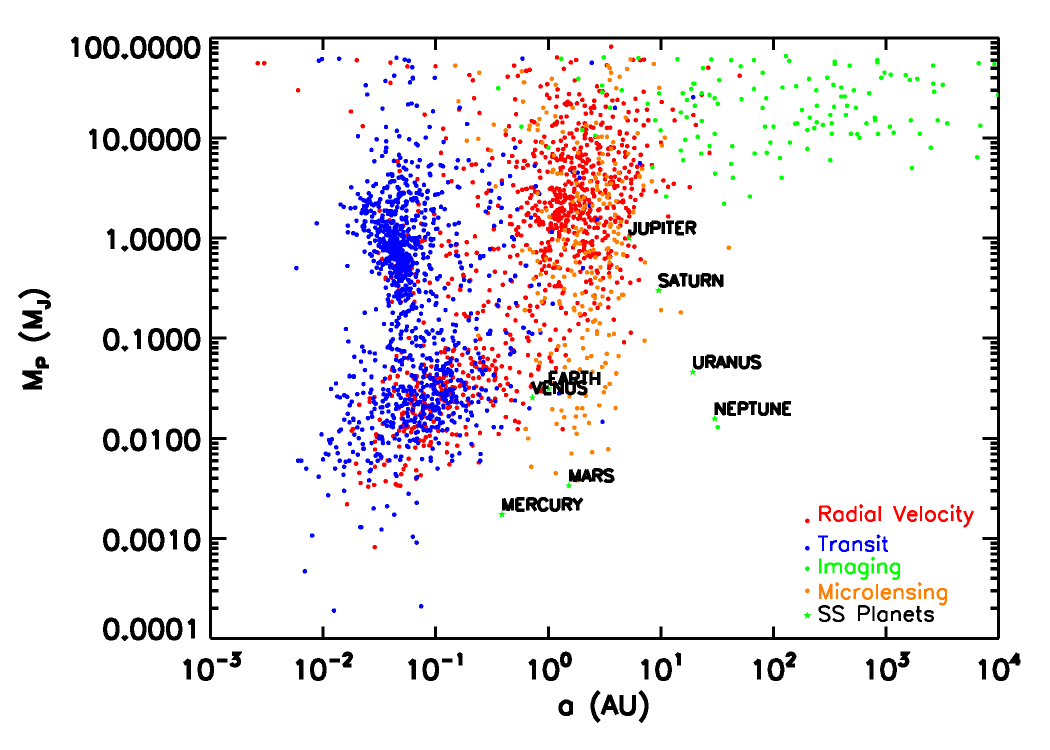}
\caption{The distribution of the masses of exoplanets discovered so far as a function of the orbital separation. The~different colors identify the different methods by which the planets have been discovered. The~planets of the Solar Systems are also reported. Data are from \url{http://exoplanet.eu/}, accessed on 31 August 2024. \label{fig:fig1}}
\end{figure}

The high difference in luminosity and the proximity of the fainter object are the two critical elements involved in high-contrast imaging, which this technique needs to achieve in astronomical observations. In~fact, directly observing very faint companions (planets or brown dwarfs) of bright stars is one of the most challenging tasks because of their small angular separation (e.g., <500 mas for a 5 au orbital radius at 10 pc) and the flux contrast between the two. {For example, for~the Solar System at a distance of 10 pc, the~contrast can range from $10^{-3}$ for Jupiter in the mid-infrared to $10^{-10}$ for Earth in the visible spectrum (e.g., \citep{traub2003esasp539_231, kaltenegger2017araa55_433})}. 
These difficulties pushed the use of indirect methods in searching for faint companions around stars. In~fact, up~to now, most of the extrasolar planets detected (about 7300 confirmed objects in August 2024, see Figure\ \ref{fig:fig1}) have been obtained by searching for the dynamical and photometric effects that invisible companions have on the status of its host star. The~most efficient indirect methods are transit (more than 6000 planets) and radial velocity (more than 1500) measurements\endnote{\url{http://exoplanet.eu/}, accessed on 31 August 2024.}. Nevertheless, these two techniques are biased toward planets in close orbits. There are about sixty planets discovered with the radial velocity and just one, HD 75784 c, with the transit method, out of the 212 known planets with their orbital axis $>5$\ au.
On the contrary, direct imaging is biased toward the large orbital axes with four detections of PMCs at a separation of $< 5$ au\endnote{\url{https://exoplanetarchive.ipac.caltech.edu/index.html}, accessed on 31 August 2024.}: WISE\,J033605.05-014350.4, 
 a~system with two BDs observed with JWST \citep{calissendorffetal2023apj947_l30}; CFBDSIR\,J145829+101343\,B, a~component of another BD binary system observed with Keck laser guide star adaptive optics \mbox{imaging \citep{liuetal2011apj740_108};} \mbox{Luhman 16 A B,} a~BD discovered with joint efforts of imaging, astrometry, and radial velocity \mbox{techniques \citep{garciaetal2017apj846_97, bedinetal2024an34530158};} and~HD\.206893\,c , which was also detected in this case with joint use of imaging, astrometry, optical interferometry, and~radial velocity techniques \citep{hinkleyetal2023aa671_L5}.
Direct imaging techniques with space-based telescopes and large ground-based telescopes in combination with adaptive optics (AO) modules is particularly well-suited for wider orbits, beyond~the snow line, and~younger (brighter) low-mass companions~\cite{lafreniereetal2007apj_670_1367,chauvinetal2010aa_509_A52,macintoshetal2015sci350_64, desideraetal2021aa651_a70, langloisetal2021aa651_a71, viganetal2021aa651_a72}. 

Direct imaging, or~the direct detection of photons from the planet and/or disk picked through the glare of a host star, allows us 
to analyze the light reflected by the planet (visible) and, in addition, the intrinsic radiation emitted by the planet itself (infrared), by exploiting colors (by means of photometric filters) and spectra (by means of conventional or integral field spectrographs) independently of their orbital inclination and, above all, independently of the activity of the parent star. Thus, it is possible to obtain information about the physical and chemical characteristics of a planetary atmosphere without waiting for the transit (transmission spectroscopy) (e.g., \citep{brown2001apj553_1006,tinettietal2007apj654L_99}) of the planet or its occultation (emission spectroscopy) (e.g., \citep{charbonneauetal2005apj626_523}).
The study of multi-band photometric images and spectra in both the visible and infrared makes it possible to estimate, under~a variety of circumstances such as distance from the star, age, etc., the~orbital and physical parameters of the planet, the~structure and the composition of its atmosphere, its surface properties, the rotation rate, and~the possible presence of life on that planet. The~planetary mass is the only parameter that the direct methods cannot measure, and generally, it is also inferred by theoretical evolutionary models (e.g., \cite{fortneyetal2008apj683_1104, burrowsetal2010apj719_341, baraffeetal2008aa482_315}) if it is subject to unknown initial conditions at very young ages. Large differences in the inferred value of planetary mass are foreseen between the so-called `hot-start' and `cold-start' models~\cite{marleyetal2007apj655_541, spiegelandburrows2012apj745_17S}.

{If an observer at distance $d$(pc) observes a star with a planet at projected distance $\delta$ (au) from its host star, then the angular separation between the planet and the star is $\theta=\delta/d$, where $\theta$ is expressed in arcsec}. This quantity gives us an idea about the angular resolution that a telescope of diameter $D$ should have in order to resolve any faint companion. A~perfect, diffraction-limited imager can discern two sources when they are separated by at least $\lambda/D$ in angle~\cite{oppenheimerandhinkley2009araa47_253}. 
So to resolve a planet that orbits its host star at 5 au at a distance from the observer of 10 pc, the~telescope should have $\theta \leq 500$\ mas, which is achievable with a 4 m class telescope {in the visible band}. However, if~we would like to see a further star as close to the star as 1au, 
say 50 pc, we need $\theta \leq 20 $\ mas. Considering that most of the faint companions (e.g., extrasolar planets) are so cool that we need infrared bands to observe them, the~problem becomes {more difficult}. Reaching such a high angular resolution is a necessary but insufficient condition. As~a matter of fact, we have to fight against the glaring star and other difficulties, too. 
To achieve high angular resolution, two main techniques can be used: single-pupil observations with large telescopes working close to their diffraction limit and~interferometric (or multiple pupil) observations that coherently combine the light from a set of individual telescopes (two units or more) with ground separations up to a few hundred meters apart. The~former case achieves typical angular resolutions of 50\,mas (i.e., 5\,au at 100\,pc), while the latter reaches angular resolutions down to  $\sim$1 mas.

The single pupil technique uses a single ground or space-based telescope with a large primary mirror. Most of them (e.g., Palomar, CFHT, Keck, Gemini, Subaru, VLT \citep{daviesandkasper2012araa50_305}) are equipped with adaptive-optics modules coupled with coronagraphic systems {to suppress or reduce the halo of the star image} and can, thus, achieve high-contrast~imaging. 
 
 Stellar interferometry achieves an angular resolution equal to $\lambda/2B$, with $B$ being the separation between the telescopes, which is referred to as `baseline'  (see \citep{monnier2003rpph66_789} and references therein). With~baselines of about a few hundred meters, interferometry reaches resolving power equivalent to that of single telescopes larger than even the most ambitious Extremely Large Telescope projects considered so far \citep{absilandmawet2010aarev18_317}.
With its angular resolution of typically 1 mas in the near-infrared on hecto-metric baselines, stellar interferometry is the tool of choice for investigating the innermost parts of circumstellar discs in nearby star-forming regions. In~particular, interferometry is currently the only suitable method for directly characterizing the most important region of protoplanetary discs where dusty grains sublimate and where accretion/ejection processes originate. 
{In principle}, interferometry also provides a sharp view of the region where terrestrial planets are supposed to be formed~\cite{absilandmawet2010aarev18_317}. 
For more evolved planetary systems, interferometry can be used to constrain the presence of circumstellar material in the inner few au, including large amounts of warm dust, (sub-) stellar companions, or even hot planetary-mass companions. In~this review, we limit ourselves to the description of the single pupil systems and their main results (see Section\ \ref{sec:res}) due to limited space. Readers interested in optical interferometry can take into consideration the following reviews \citep{quirrenbach2001araa39_353, saha2002rvmp74_551, absilandmawet2010aarev18_317, gravitycollaboration2017aa602_A94, defrereetal2020SPIE11446E_1J}. 

This review has a mainly didactical aim, and it is limited to the description of the principal difficulties in direct imaging observations and the technological solutions for overcoming them (Section\ \ref{sec:dirmethiss}), and~the observing strategies and techniques for suppressing speckle noise (Section\ \ref{sec:HCI}). Furthermore, we describe the present and future instrumentation that allows high-contrast imaging observations (Section\ \ref{sec:instr}) and the typical algorithms used for post-processing analysis of data (Section\ \ref{sec:sdec}). In~Section\ \ref{sec:res}, we discuss some results with the description of some remarkable objects (at the personal evaluation of the authors), and in Section\ \ref{sec:statistics}, we describe the surveys dedicated to the search for planets around young stars and their occurrence rate. In~Section\ \ref{sec:conc}, we outline perspectives and conclusions. Direct imaging and high-contrast imaging have been described in several reviews (e.g., \citep{oppenheimerandhinkley2009araa47_253, traubandoppenheimer2010exopbook_111, bowler2016pasp128_j2001, currieetal2023ASPC534_799, follette2023pasp135_3001}) to which the reader is referred for further information.

\section{Direct Observing Issues, Difficulties, and~Solutions}
\label{sec:dirmethiss}
\begin{figure}[H]
\includegraphics[scale=.33]{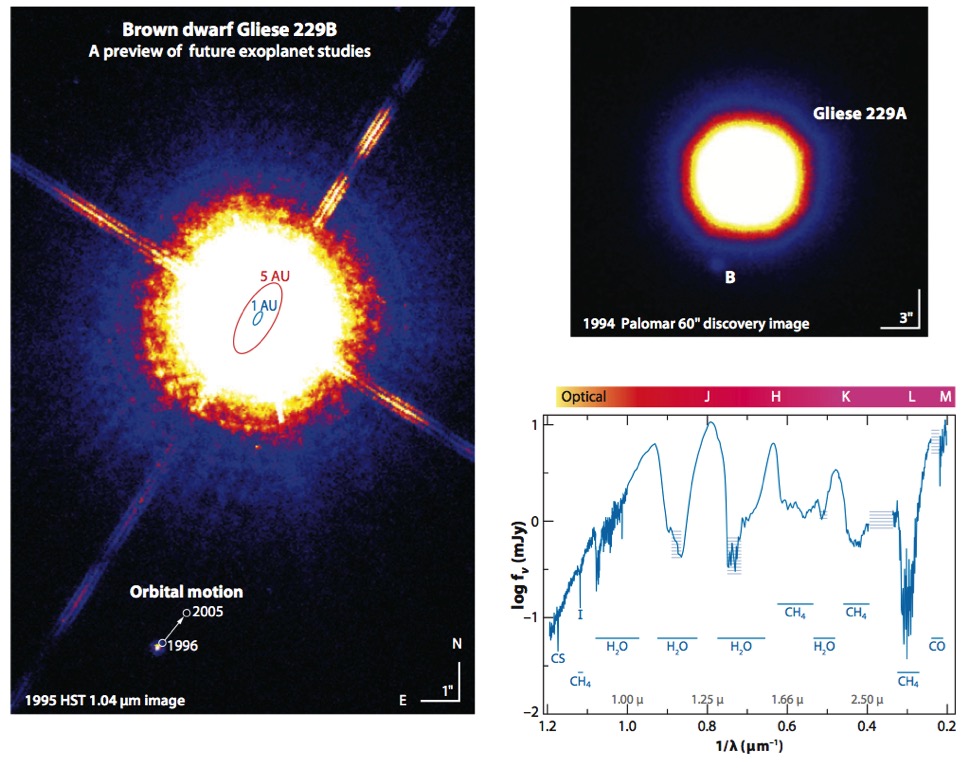}
\caption{HST and Palomar images of the Gliese 229 system. The~M-dwarf, and~T-dwarf pair has been discovered with coronagraphy. {Gliese 229 B is at a projected distance  $>$ 7 arcsec from its host star, and the luminosity contrast between the two objects is about $10^{-4}$ in the optical and NIR. This is just an example of the potentiality of DI observations.  Both the limited projected separation and the high contrast made the star overcome the light of the companion; see the 1\,au and 5\,au orbits inside the glare of the star}. Left: the Gliese 229 system observed with the HST. Right---up: the Gliese 229 system observed with Palomar. Right---down: the Gliese 229 B spectrum. The~Picture was taken by \citet{oppenheimerandhinkley2009araa47_253}.}
\label{fig:5}       
\end{figure}

Figure\ \ref{fig:5} shows 
the potentiality of the direct imaging methods.
In this case, a~basic stellar coronagraph and a standard spectrograph were sufficient to acquire the architecture of the system, and~the spectroscopy of the low mass companion~\cite{oppenheimeretal2001aj121_2189}.

%

This example is a favorable  case that is not so common if we search for planets with DI due, as~already told in the previous section, to~the small projected separation and to the luminosity contrast.
The former depends on the evolution history of the system and it could reach a few tenths of arc-seconds for planets at a few au from stars at a distance up to 100 pc from the Sun. 
The contrast is a function of the wavelength $\lambda$, the~properties of the planet, the~apparent geometry, and~the age of the system:
\vspace{6pt}
\begin{equation}
\begin{array}{c}
\text{C}_{\text{OPT}}=\frac{\text{F}_{\text{p, Reflected}}(\lambda)}{\text{F}_\star (\lambda)};\\
\\
\text{C}_{\text{IR}}=\frac{\text{F}_{\text{p, Intrinsic}}(\lambda)}{\text{F}_{\star} (\lambda)}.\\
\end{array}
\end{equation}
\noindent
where $\text{C}_{\text{OPT}}$ and $\text{C}_{\text{IR}}$ are the contrast in the optical due to the reflected flux ($\text{F}_{\text{p, Reflected}} (\lambda)$) from the planet and the contrast in the IR due to the thermal or intrinsic emission ($\text{F}_{\text{p, Intrinsic}} (\lambda)$) of the planet respectively.   
$\text{F}_\star$ is the flux emitted by the host star.
The reflected component depends on the brightness of the host star, the~separation of the planet by the star, and~the albedo $\text{A}(\lambda, t)$ of the planet, which depends on the wavelength and time $t$. The~intrinsic component is defined by the temperature T$_\text{p}$ of the planet itself. 

Considering a planet with a radius $\text{R}_{\rm p}$ that orbits the host star at an orbital distance a, the~flux reflected by the planets is as follows:\vspace{-6pt}
\begin{equation}
\text{F}_{\text{p, Reflected}}=\text{A}(\lambda,t)\phi(t)\frac{\text{R}^2_{\rm p}}{4\text{a}^2}\text{B}(\lambda,\teff)\text{R}_\star^2,
\end{equation}

\noindent
where $\text{B}(\lambda, \teff)$ is the brightness of the host star and $R_\star$ is the stellar radius. $\phi(t)$ is the phase angle ([$0,\pi]$) or the angle between the observer--planet and the planet--star direction. Due to the irradiation characteristics of the host stars, the~flux reflected by the planet is in the visible {range}.  The~albedo $\text{A}(\lambda, t)$ represents the fraction of the stellar light reflected by the planets, and~it depends on what properties of matter are taken into consideration. Both geometrical ($\text{A}_\text{g}$) and Bond albedo ($\text{A}_\text{B}$) are considered to describe the general reflectance properties of a planetary surface. The~geometrical albedo is the ratio of planet brightness at phase angle $\phi = 0$ (superior conjunction) to the brightness of a perfectly diffusing disk with the same position and apparent size as the planet. On~the other hand, the~Bond albedo is the ratio of the light reflected by the planet to the light incident to the planet,  which is integrated over all wavelengths and~on all the phase angles. The~Bond and geometric albedo are correlated by means of the phase integral q: $\text{A}_\text{B}=\text{A}_\text{g}\text{q}$.

We can determine the intrinsic component of the flux emitted by a planet as~that of a black body at the same temperature (T$_\text{p}$) as the planet itself:
\begin{equation}
\text{F}_{\text{p, Intrinsic}}=\epsilon_{\text{IR}}\text{B}(\lambda, \text{T}_{\text{p}})\text{R}^2_{\text{p}}
\end{equation}
where $\epsilon_{\text{IR}}$ is the emission coefficient of the planet, and it depends on the wavelength and on the chemistry of the planetary surface and atmosphere. The~value of the T$_\text{p}$ results from the absorbed fraction of the host star radiation plus the internal heat of the planet. If~the latter is negligible, the planetary temperature is named equilibrium temperature (T$_{\text{eq}}$). If this is not the case, it is named the effective temperature of the planet ($\teffpl$) instead. In~addition, the~atmosphere of the planet could be a source of heat by means of the greenhouse effect, and~generally, its contribution is added to the equilibrium or the effective temperature of the planet. In~the Earth's case, for example, the greenhouse effect raises the temperature by about 30 degrees~\cite{kastingandcatling2003araa41_429}. The~fraction of the stellar radiation absorbed by the planet is $(1-\text{A}_{\text{B}})$, and~assuming that the incident flux is equal to the radiated flux:\vspace{6pt}
\begin{equation}
\text{T}_{\text{eq}}=\left(\frac{1-\text{A}_{\text{B}}}{4f}\right)^{1/4}\left(\frac{\text{R}_{\text{p}}}{\text{a}}\right)^{1/2} \text{T}_\star.
\end{equation}

$f$ is the fraction of the area over which the heat is spread, so it is $f = 1$ for a rapid rotator and $f = 0.5$ for a tidally locked planet. The~$\teffpl$ results from fitting the intrinsic emission with a black body curve. The~two temperatures are generally different from each other,  for~example in the case of Jupiter: $T_{\rm eq}=110\ K$ and $\teffpl=124.4\ K$ \cite{depaterandlissauer2010plscbook}).

\begin{figure}[H]
\includegraphics[scale=.20]{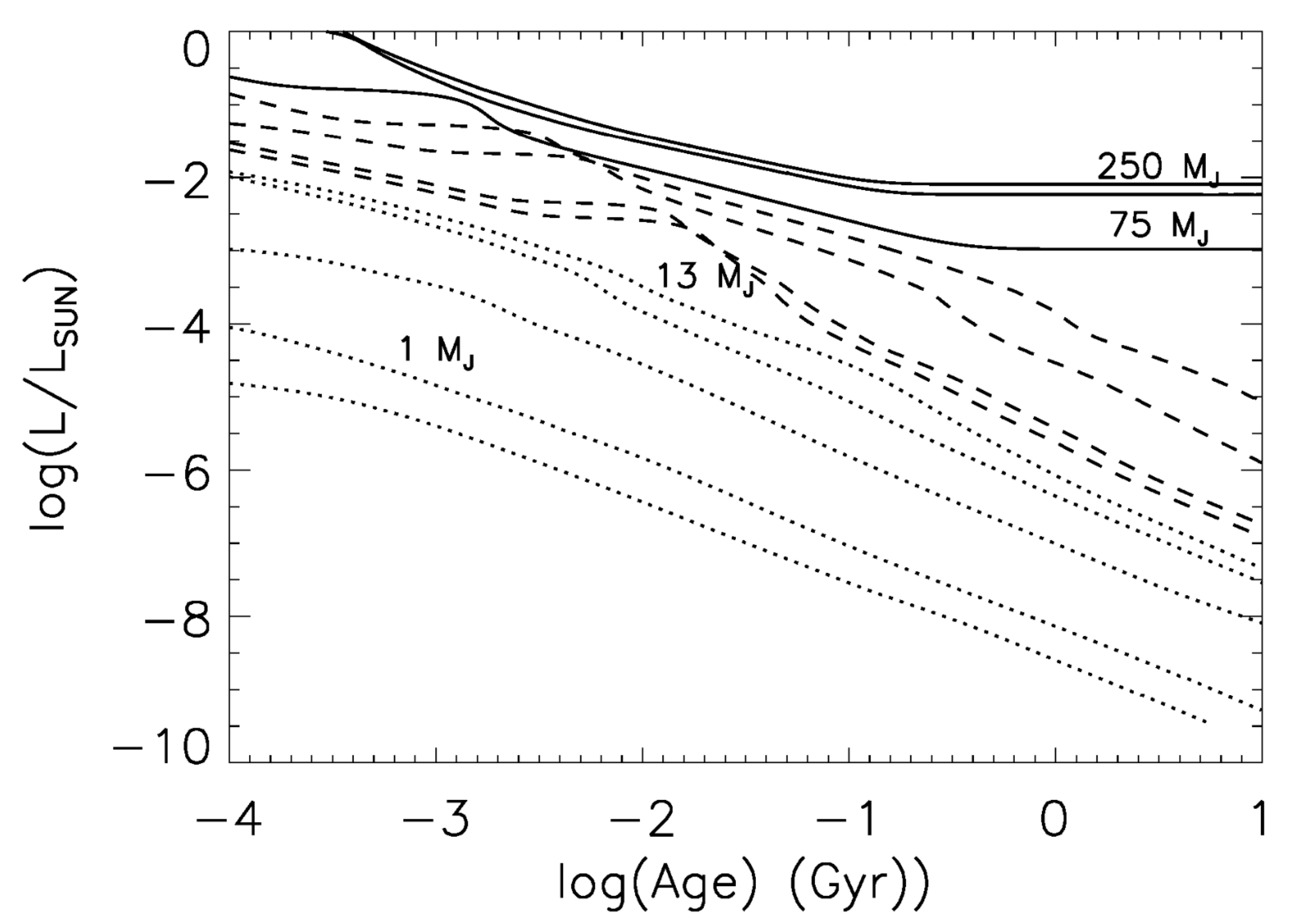}
\caption{Theoretical models for the luminosity evolution of different structures with different masses versus age.  The~stars are shown in (continuos line), while sub-stellar structures with M > 13 $\text{M}_\text{J}$ are in (dashed line), and~giant planets are in (dotted line). The~masses of the structures are labeled in Jupiter mass units. Young planets are brighter by more than three orders of magnitude than old planets. The~data are taken by \citet{burrowsetal2001rvmp73_719}. }
\label{fig:theoretical}       
\end{figure}

The expected contrast of the typical Jupiter-like measurements (for an angular separation of \mbox{0.5 arcsec}) is of the order of $10^{-9}$ in the visible band and $10^{-6}$ in the NIR. On~the contrary, for~Earth-like planets orbiting the habitable zone of a G star (at angular separations of about 0.1\,arcsec), the~contrast is $10^{-10}$ and $10^{-7}$ in the visible and NIR, respectively. {For further details, the interested reader is referred to \citep{traubandoppenheimer2010exopbook_111, currieetal2023ASPC534_799}}.
Young planets could also be three orders of magnitude brighter than old ones (Figure\ \ref{fig:theoretical}, \cite{burrowsetal2001rvmp73_719, baraffeetal2003aa402_701}).  This reflects on the observational strategy (see ahead) because the contrast depends on the distance from the star for old planets with negligible intrinsic flux component,  while it is independent of the distance for young planets that have high intrinsic flux. A~detailed spectroscopic study of even young ($\sim100$\,Myr) and hot ($\sim 800$\, K) planets of about 1 M$_{\rm J}$ requires a contrast of $10^{-8}$. It depends on the distance by the host star and on the age of the system, but~also if it is possible to detect one photon from the planet for every $10^6$ from the star, to~probe the depths of the spectral features requires at least two more orders of magnitude in precision. The~contrast is also worse for old planets, making this kind of study really challenging. In~the case of Terrestrial planets, it becomes worse, requiring a contrast of $10^{-10}$ \cite{oppenheimerandhinkley2009araa47_253, traubandoppenheimer2010exopbook_111}.

\vspace{-6pt}

\subsection{Image and Speckles~Formation}
\label{sec:speck}
The light coming from a star far away from the observer, once it reaches the observer's telescope, could be considered as a plane wavefront light. For~example, for~light coming from a 10 pc distant star, the~deviation from a plane is of one part on $10^{17}$ \cite{oppenheimerandhinkley2009araa47_253}. Following Huygens' Principle, the~telescope aperture D becomes an emission center of spherical waves that will focalize onto the telescope focal plane. In~this case, considering a circular aperture D and~a wavefront with an amplitude equal to $\text{A}_1(\text{X})=1$, we have the following on~the focal plane of the telescope (Chapter 8.5.2, \cite{bornandwolf1999propbook}):
\begin{equation}
\text{A}_2(\theta)=\sqrt{\text{I}_0}\frac{2\text{J}_1(\pi \text{D} \theta/ \lambda)}{\pi \text{D} \theta/ \lambda} ,
\end{equation}

\noindent
and the corresponding intensity is $\text{I}_2 = |\text{A}_2|^2$, or~\begin{equation}
\text{I}_2(\theta)=\text{I}_0\left[\frac{2\text{J}_1(\pi \text{D} \theta/ \lambda)}{\pi \text{D} \theta/ \lambda}\right]^2 .
\end{equation}

\noindent
$\text{J}_1(\text{X})$ is the Bessel function of the first order. The~first angle at which the intensity is zero is $\theta_z=1.22 \lambda /\text{D}$ or the Airy disk radius. The~full width at half-maximum (FWHM) of the intensity pattern is $\theta_z=1.03 \lambda /\text{D}\cong \lambda/D$. This value is often mentioned as the diameter of the diffraction-limited image of a point source of a telescope of diameter D. For~a ground-based telescope, the light from a star passes through about 20\ km of atmosphere, and~the image of the point source is degraded to $\lambda/\text{r}_0$, where $\text{r}_0$ is the Fried parameter (see \mbox{Section\ \ref{sec:AO}}), the~indicative size of the seeing cell size, typically about 10 cm in the visible band ( $\text{r}_0 \propto \lambda^{6/5}$). In~an atmospheric cell of radius  $\text{r}_0$, the~refraction index of air is constant but~different from that of adjacent atmospheric cells. The~incoming wavefront is heavily modified, and~the point source image on the telescope's focal plane is degraded (seeing image). To~solve the problem of seeing there are two solutions. The~first is completely bypassing the Earth's atmosphere (e.g., HST, JWST, etc.). The~second solution is to equip the telescope with an adaptive optics (AO) module to correct the atmospheric turbulence effect on the wavefront in real time and restore almost the full resolution of the telescope (see Section\ \ref{sec:AO}).

\begin{figure}[H]
\includegraphics[scale=.30]{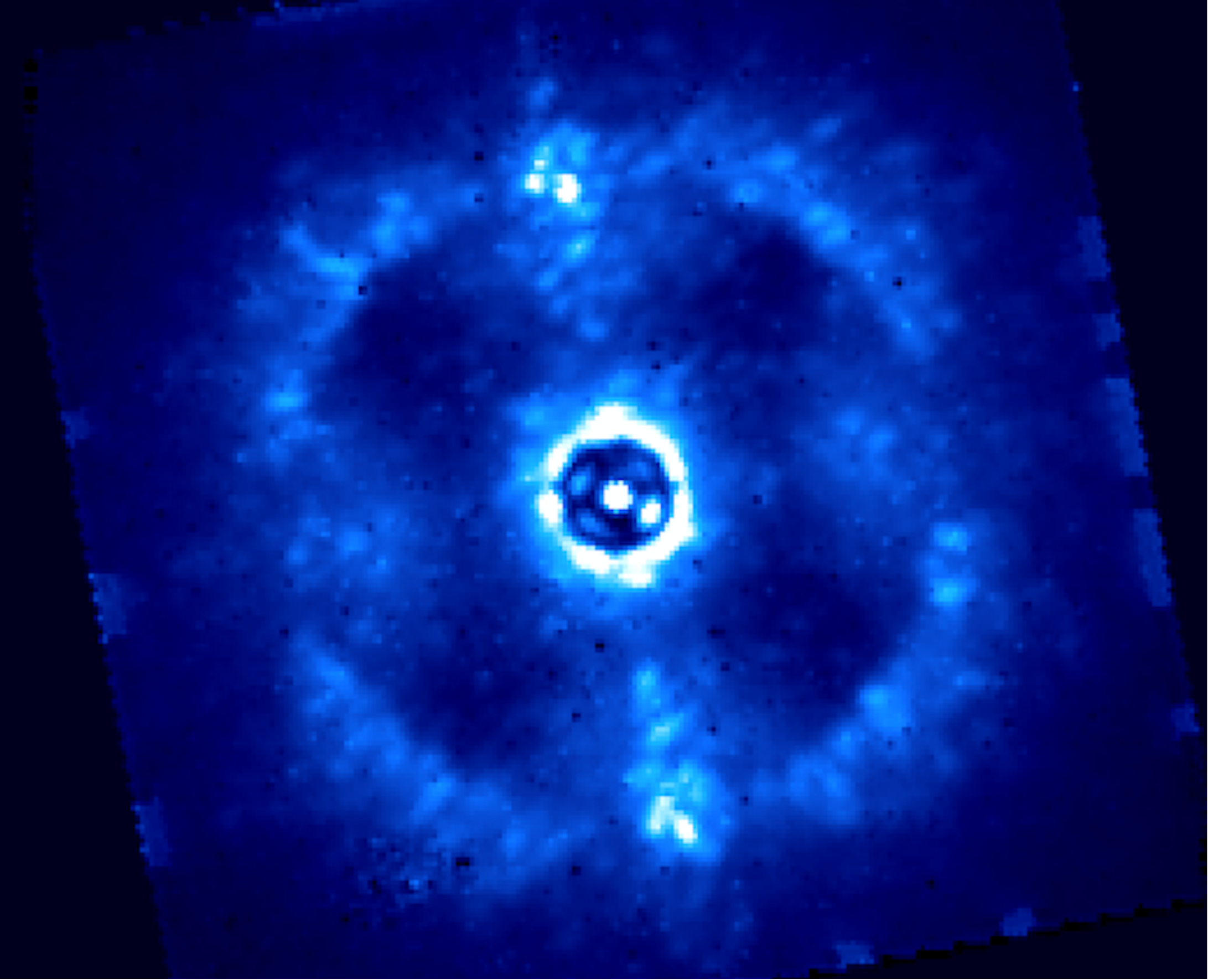}
\caption{Image taken with a coronagraph showing the presence of speckles.  
}
\label{fig:9}       
\end{figure}

The image obtained using a short exposure time to image a point source without any wavefront correction is composed of {an interference pattern with a short lifetime named speckles} (see Figure\ \ref{fig:9}). Speckles are interference patterns due to several coherent patches of size $r_0$ \cite{reacineetal1999pasp111_587} that are distributed above the aperture (D) of the telescope. One aperture (sub-pupil) of dimension $r_0$ produces a Point Spread Function (PSF) with a FWHM$\sim \lambda/r_0$. Two such sub-pupils at a distance of $\sim D$ from each other constitute an interferometer that generates on the focal plane of the telescope an interference pattern of size $\sim\ \lambda/D$ with fringes that run orthogonally to the conjunction of the two apertures. 
Each couple of apertures causes a different system of interference fringes, and~when the interference is a constructive one, it causes a bright speckle. The~random phase variation of an incoming wavefront enables the movement of the interference pattern inside the dimension of the PSF, which causes the characteristic boiling of speckles and makes the intensity of speckles vary in a random way. The~wavelength of the observed light and the wavefront perturbations determine the position of the speckle on the focal~plane.

The fluctuations of the atmospheric diffraction index are carried along by the wind, and~the timescale in which the wavefront changes depends on the wind speed in the overlying atmosphere. Typically, we have a speckle lifetime of 3 ms for a wind speed of 10 m/s. The~speckle lifetime depends on the telescope site, the~AO systems, and~ancillary instrument configurations. Therefore, a~ground-based telescope image of a star is formed of approximately $(D/r_0)^2$ speckles, churning on a timescale of the order of milliseconds, and~spread over an angular diameter on the sky of  $\sim \lambda/r_0$ or about 1\ arcsec in the visible, independent of telescope diameter. Also, the~telescope and instrument optics induce speckles generated by the optics surface shape errors due mainly to the manufacturing~processes. 

%

Generally, the~behavior of the speckles does not follow the Poisson statistics, and~they represent a noise several orders of magnitude larger than the shot noise of a perfect PSF~\cite{reacineetal1999pasp111_587}. They are evanescent and exhibit a correlated noise behavior~\cite{soummeretal2007apj669_642}. These characteristics do not allow blurring them, enhancing the exposure time, or~using broadband observations to improve the system sensitivity to objects fainter than the speckle's background. Nevertheless, several techniques or observation strategies exist that can remove or reduce the barrier of~speckles.

\subsection{Adaptive~Optics}
\label{sec:AO}
\begin{figure}[H]
\includegraphics[scale=.50]{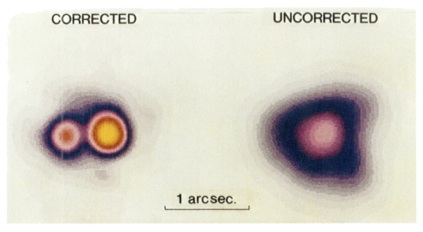}
\caption{First image of the AO prototype `COME--ON'  system taken at 1.52 m telescope of the Observatoire de Haute Provence. $\gamma_2$\ And (a binary star with a 0.5$''$ separation) was observed in $K$ band~\cite{rousset1990aa230_L29}. }
\label{fig:10}       
\end{figure}

In 1953, the astronomer Horace W. Babcock wrote: ``\emph{If we had the means of continually measuring the deviation of rays from all parts of the mirror, and~of amplifying and feeding back this information so as to correct locally the figure of the mirror in response to the Schlieren pattern, we could expect to compensate both for the seeing and for any inherent imperfection of the optical figure
}'' (\cite{babcock1953pasp65_229}). The~realization of Babcock's concept had to wait until the assembly of the first astronomical AO instrument COME-ON tested at the end of 1989 at the 1.5 m telescope of the Observatoire de Haute-Provence (see Figure\ \ref{fig:10}), and then it was installed on the 3.6 m ESO telescope in La Silla~\cite{merkleetal1989msngr58_1, rigautetal1991aa250_280, beckers1993araa31_13}.

%

The main elements of an AO system are the wavefront sensor (WFS), a~real-time computer, and~a deformable mirror (DM). A~WFS allows us to measure the wavefront, providing a signal that lets us estimate the shape of the wavefront with sufficient accuracy. It generally includes a phase-sensitive optical device and a low-noise, high-quantum efficiency photon detector. Among~the different WFSs, the~most used is the Shack-Hartmann WFS. It is a simple system that is able to sample a transferred pupil with a large number of small lenses. Each lenslet forms an image of the star onto multiple quad cells or a CCD, and~the position of each image caused by the local tilt of the sampled portion of the wavefront could be measured by obtaining information on the average slope of the wavefront. 
Another very used type of WFS is the Pyramid one, which adopts a pyramidal prism with a very large apex angle to split the incoming beam into four images that will be perfectly identical in the absence of any aberration. Anyhow, any aberration in the wavefront will introduce differences in the four pupil intensity distribution that can be used to obtain information about the aberration itself \citep{Ragazzoni1996JMOp43_289}.
This information can eventually be used to drive a DM, and~a closed-loop control is established. 
The role of a DM is to correct for the optical path differences introduced by the turbulent atmosphere. A~DM comprises an array of actuators connected to a thin reflecting surface deformed by the expansion of the actuators. The~most important parameters for a DM are stroke, response time, spacing, and~the number of actuators. There are some telescopes that mount the DM directly to their secondary mirrors (DMS), like, for example, the 6.5\ m Multiple Mirror Telescope (MMT) and the Large Binocular Telescope (LBT) (\cite{espositoetal2010spie7736E_09}), but~typically DMs are medium-sized piezo DMs with an actuator spacing of several millimeters. SPHERE~\cite{beuzitetal2019aa631_A155}, the~high-contrast imager of VLT, mounts this type of DM. Recently, MOEMSs (micro-optical-electrical-mechanical systems) have emerged as an alternative to DMs. They are manufactured using standard semiconductor technologies. The~Gemini Planet imager (GPI \cite{macintoshetal2018spie10703E_0K}) at the Gemini South telescope used them (GPI is under refurbishment; see Section\ \ref{sec:instr}).

\begin{figure}[H]
\includegraphics[scale=.30]{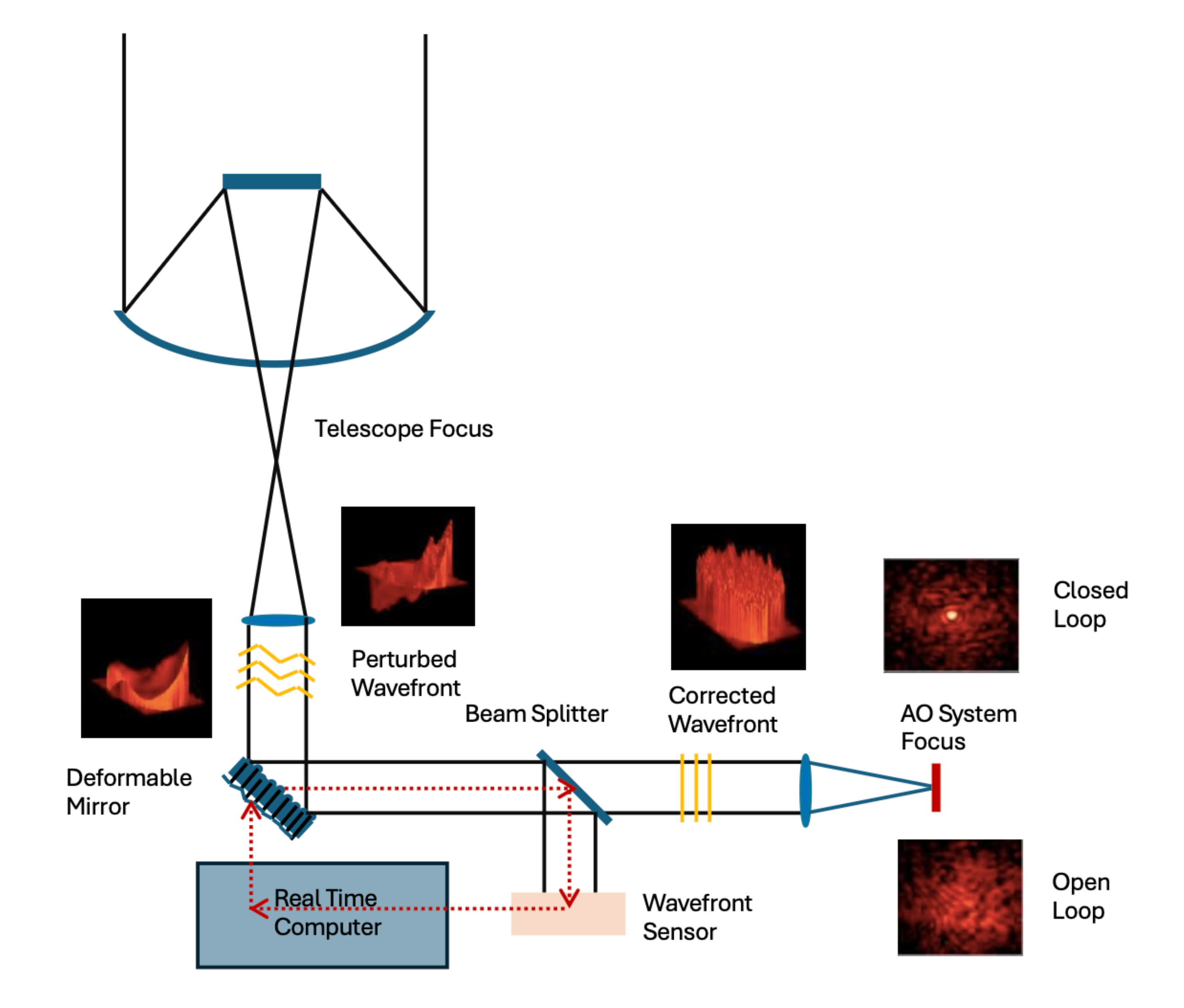}
\caption{The principle of an  Adaptive Optics System. The inserted images represent the different status of the wavefront: before the closure (perturbed wavefront) and after the closure of the AO control loop (corrected wavefront). 
}
\label{fig:12}       
\end{figure}

The primary application of AO is the real-time correction of the image seen on the focal plane of a telescope to produce a diffraction-limited image. This result is achieved in several steps, the~first of which is obtained with a tip/tilt system, or~finer guidance tracker, able to stabilize the stellar PSF against the large movements (up to a few arcsec) due to the atmospheric variation and instability, and~telescope vibrations. The~remaining alterations of the incoming wavefront are adjusted by a DM. The~AO system regulates the optical path variations (wavefront), measures the deviations from a plane wavefront with a WFS, and~uses a real-time computer; it calculates and applies the appropriate correction to a DM that reaches a shape able to correct the input wavefront to a quasi-plane wavefront (see Figure\ \ref{fig:12}).

The AO system performs a feedback loop that is carried out hundreds of times per second to comply with the time scale of the atmospheric changes. The~dimension of the resolution elements of the WFS (sub-apertures) and the number of the DM active sub-surfaces (actuators) projected on the telescope entrance aperture should approximately match the $r_0$.  The~efficiency of an AO system is measured by the Strehl ratio or the ratio of the peak intensity in a real image to that of a perfect image taken with the same imaging system working at the diffraction limit. Named $\sigma_{\text{WFE}}$  the root mean square of the wavefront error in radians, the~Strehl ratio could be expressed by the Mar\'echal approximation $\text{S} \sim \text{exp}(-2 \pi \sigma_{\text{WFE}}^2)$  when $\sigma_{\text{WFE}} \ll1$ \citep{ross2009apopt48_1812}. {The Strehl ratio depends on the wavelength at which the imaging system is working, or~in other words, it depends on the photometric band used in the observation (e.g., \citep{daviesandkasper2012araa50_305}}).

We have just described single-conjugate adaptive optics (SCAO), in~which the wavefront coming from a single natural guide star (NGS) is measured.  This is the technique generally used in HCI observations. Nevertheless, other than SCAO, there are other AO systems to optimize the use of the guiding star exploiting synthetic stars generated by laser beacons or laser guide stars (LGSs). 
This  LGS technology exploits two physical mechanisms: the Rayleigh scattering in the dense regions of the atmosphere up to altitudes of about 30 km and~the resonance fluorescence of sodium atoms concentrated in a layer at about 90 km height. Several telescopes around the world are now equipped with AO modules that use LGS, like, for~example, Keck II \citep{wizinowichetal2006pasp118_297}, VLT \citep{bonaccinicaliaetal2006spie6272E_07}, Gemini North \citep{boccasetal2006spie6272E_3LB}, and~Subaru \citep{hayanoetal2010spie7736E_0NH}. 
 Other AO systems are also designed to probe different turbulent atmospheric layers with different combinations of NGSs or LGS (we just mention them inviting the interested reader to the review by \citet{daviesandkasper2012araa50_305}): the Laser Tomography Adaptive Optics (LTAO), Ground Layer Adaptive Optics (GLAO), and~Multi Conjugate Adaptive Optics (MCAO).

In the last decade, most of the new generation high-contrast imagers adopted the extreme AO (ExAO) concept. The~ExAO system is built to fully correct both atmospheric and instrumental perturbations using a single star on-axis for wavefront measurement. It is not different from conventional SCAO but~challenging in the technical implementation; the main aim is to provide exceptionally high performance on bright ($V<10$\ mag) stars, reaching a Strehl ratio of about 90\% in the $H$-band~\cite{guyon2018araa56_315}. These results have been obtained by optimizing the sampling of the DMs, enhancing the number of actuators, and using fast and low noise detectors such as Electron-Multiplying Charge-Coupled Devices (EMCCDs) to reduce the errors in the wavefront sensor measurements. 
A complete review on the ExAO is given in \citet{guyon2018araa56_315} at which the interested reader is~referred.

However, the~AO is necessary but not sufficient in performing high-contrast observations. This goal is reached with the introduction of a coronagraph in the imaging system.

\subsection{Coronagraphy}
\label{sec:cor}
\begin{figure}[H]
\includegraphics[scale=.20]{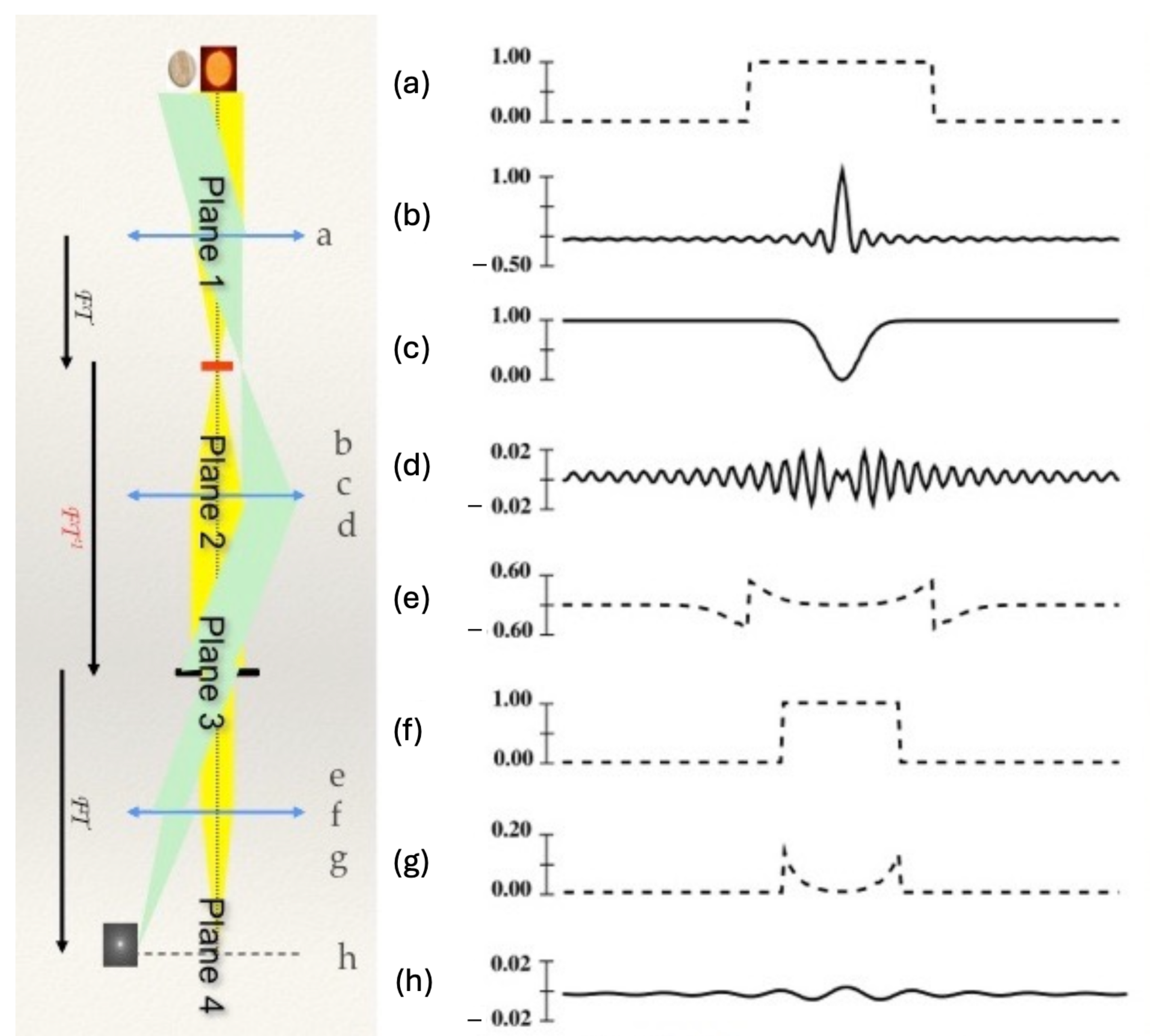}
\caption{Coronagraphy principle and Fourier Optics.The~optical scheme of a classical Lyot coronagraph (left panel). The light coming from the star (yellow, on-axis) and from the planet (light-green, off axis) are shown. On the right, the positions and electric field or stop profiles of the following: (\textbf{a}) Primary pupil for on-axis source ; (\textbf{b}) Image before image plane stop; (\textbf{c}) Image plane stop; (\textbf{d}) Image after image plane stop; (\textbf{e}) Pupil before Lyot stop; (\textbf{f}) Lyot stop; (\textbf{g}) Pupil after Lyot stop; (\textbf{h}) Final on-axis image (right panel). Modified from~\cite{sivaramakrishnanetal2001apj552_397}.}
\label{fig:lyotscheme}       
\end{figure}

Bernard Lyot~\cite{lyot1939mnras99_580}, in 1930, invented the coronagraph to study the Sun, which was a telescope with a design able to block the light coming from the solar disk to see the extremely faint emission from the corona. The~main part of this device is an occulting disk in the focal plane of a telescope completed by other optical parts to reduce stray light. After~a few years, coronagraphy has also been utilized in star observations, and~recently become an important technique for high-contrast imaging. 
In fact, large ground-based telescopes, thanks to the AO systems (see Section\ \ref{sec:AO}) obtain the diffraction image of a point source that is many orders of magnitude brighter than any faint companion which became observable after that the coronagraph suppresses the diffraction peak. The~left panel of Figure\ \ref{fig:lyotscheme} shows a scheme of a typical Lyot coronagraph. To~achieve the suppression of starlight, it uses two masks (the occulting mask c and Lyot Stop f).

 In the first stage, at~the focal plane, the~image of the target star is focalized at the center of a circular opaque mask that absorbs most of the starlight and also diffracts part of the light around it. After~that, in~the following pupil, the~residual starlight has been concentrated into a bright outer and inner ring around the conjugate location of the secondary mirror of the telescope. This light can be filtered out without greatly affecting the light from the fainter companion posing a diaphragm (named Lyot mask) here. Finally, an~image is formed by optics after this Lyot mask, and~the intensity of the host star has been reduced by about 99\%, while the companion is only affected at a level of a few percent. The~goodness and efficiency of a coronagraph are defined by four different parameters that take into account its ability to obscure the central star by affecting the smallest possible region around the star. The~main parameter is the inner working angle (IWA) that defines the passage between the obscured and the transmitted region. It is generally treated as a hard lower limit in the modeling of coronagraphic instruments, also if it is not normally a sharp function of the detection sensibility~\cite{guyonetal2006apjsuppl167_81}. Typically, IWA can range between $\sim 2 \lambda/\mbox{D}$ and $\sim 4 \lambda/\mbox{D}$. Another important parameter is the fraction of the planet's light that is not suppressed by the coronagraph. Most of them have a throughput of $\sim$80\%. The~other two parameters concern the sensitivity of the coronagraph to low-order wavefront errors (aberrations like tilt, focus, astigmatism, etc.) and the chromaticity, which is the capacity to suppress the light on a large range of wavelength: lower the chromaticity better the behavior of the~coronagraph.
 
\begin{figure}[H]
\includegraphics[scale=.40]{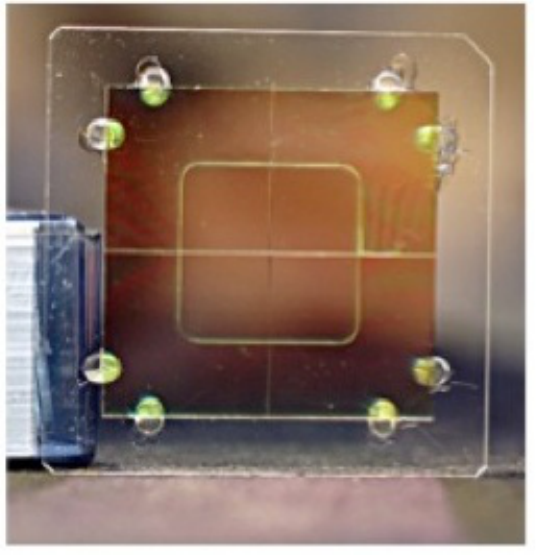}
\includegraphics[scale=.25]{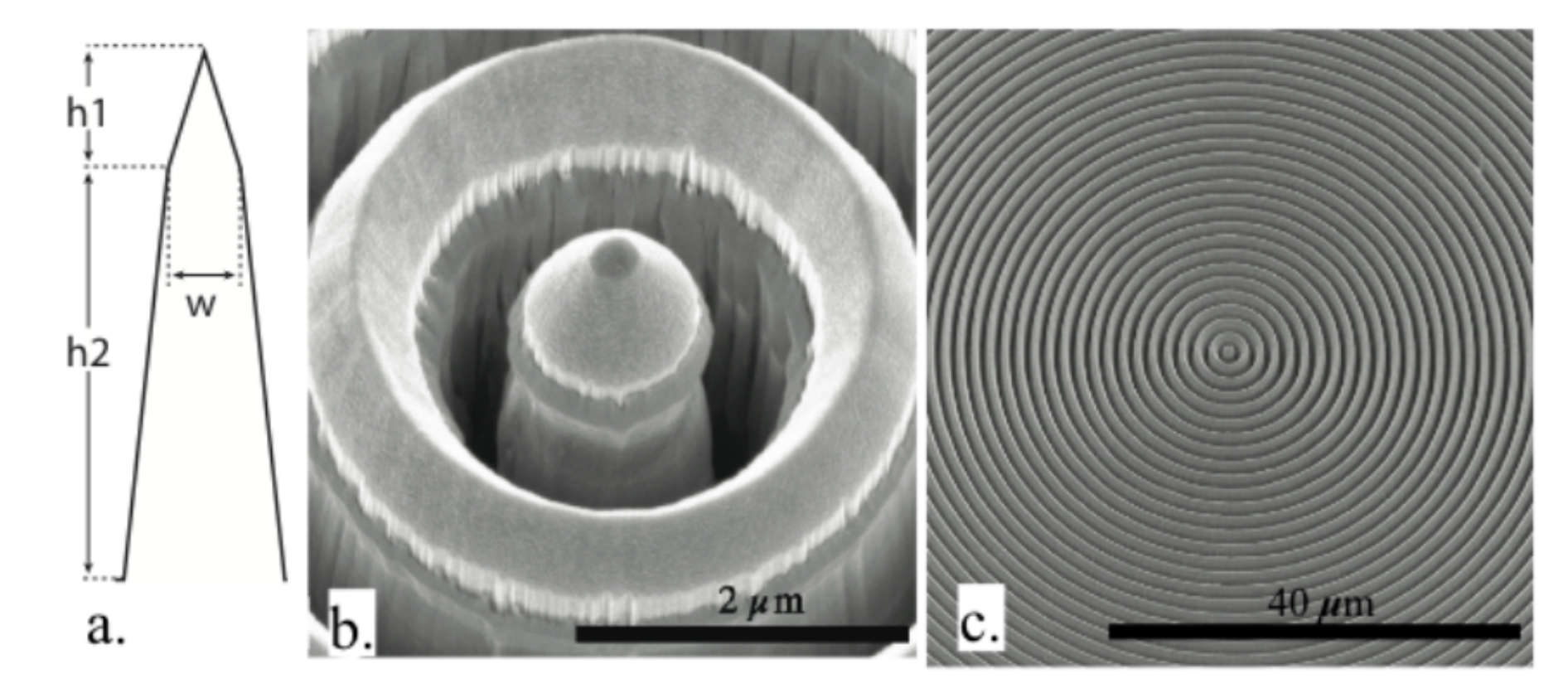}
\caption{Examples of phase focal mask for phase-based Lyot coronagraphs. Left: the 4QPM phase mask of SPHERE (described in \cite{boccalettietal2004pasp116_1061}). Right: the Annular Groove Phase Mask (AGPM) mounted on NACO at VLT.~a) Schematic view of the AGPM, b) zoom of the central part of the AGPM, c) overview of the structure of the device (for details see \cite{mawetetal2013aa552_l13}).}
\label{fig:16}       
\end{figure}

 There are at least four broad categories of coronagraphs: amplitude-based Lyot coronagraphs, phase-based Lyot coronagraphs, interferometric coronagraphs, and~pupil-apodization coronagraphs. Moreover, for each of these categories, there are a lot of different concepts. The~interested reader can refer to the~\cite{guyonetal2006apjsuppl167_81, sivaramakrishnanetal2001apj552_397, galicherandmazoyer2024crphy24_S133} reviews. Two variants of the classical Lyot coronagraph scheme (see Figure\ \ref{fig:lyotscheme}) are the band-limited coronagraph (BLC) \cite{kuchnerandtraub2002apj570_900} and~the Apodized-Pupil Lyot Coronagraph (APLC). The~main difference between the two is the manufacturing of the focal mask. The~BLC has a focal mask with a transmission profile carefully tailored to confine the residual diffracted light to a finite outer region of the pupil. Instead, APLC modifies the entrance pupil with a tapered transmission mask that allows for a reduction of the sharp discontinuity in the transmitted wavefront at the edge of the pupil, which reduces the amplitude of the secondary maxima of the diffraction figure. Those coronagraphs perform similarly, achieving a throughput of 20--40\% for an \mbox{IWA $ > 4 \lambda/D$.} Both coronagraphs are easily adapted to conventional telescopes with on-axis secondary mirrors, and~they have been selected by the major ground-based coronagraph projects. In~the Lyot class of coronagraph, there are also alternative designs that make use of phase shift focal devices to block the starlight (e.g., \cite{boccalettietal2004pasp116_1061}). In~theory, this technique removes all of the starlight and has no effect on anything else in the field of view. The~best-known Phase Based Lyot Coronagraph concept is the 4-quadrant phase-mask (4QPM) coronagraph~\cite{rouanetal2000pasp112_1479}.
The phase mask positioned in the focal plane consists of four contiguous quadrants of transparent material on which the star is focused in the center. The~adjacent quadrants differ in optical thickness by a half-wavelength (Figure\ \ref{fig:16} left panel). The~transmitted beam will be nulled on the axis, but~whatever object is focused on one of the quadrants will be transmitted.
%

A variant of the 4QPM is the Annular Groove Phase Mask (AGPM)  (see Figure\ \ref{fig:16} right panel), which consists of a spiral-staircase phase mask that generates a longitudinal phase delay by operating on both polarizations. The~center of the optical vortex is a phase singularity in an optical field, which generates a point of zero intensity~\cite{traubandoppenheimer2010exopbook_111}. These coronagraphs achieve near-theoretical performances with IWA of $1 \lambda /\mbox{D}$ and high throughput. Nevertheless, they are really sensible to the position of the image of the star in the focal plane, with performances degrading for small tip/tilt errors or partially resolved~stars.

\section{High-Contrast Imaging and Speckle~Suppression}
\label{sec:HCI}

 The main goal of high-contrast imaging is primarily to discover and characterize low-mass companions close orbiting brighter stars.   
 In the previous section, we described the main problematics and the techniques to mitigate them for direct imaging of faint companions. At~first, we need large telescopes to acquire high angular resolution. Furthermore, the~use of AO modules allows us to restore the diffraction figure of these telescopes. But~this is still not enough. In~fact, we have to eliminate the restored diffraction peak to eliminate the glare of the star, and we achieve this with a coronagraphic device. But~also now it is not enough to reach the contrast necessary to reveal the faint companion that is drowned in that variable, not smooth background, due to speckles. These image artifacts are caused by quasi-static optical aberrations within the telescope, AO system, coronagraph, or~camera that are not correctly calibrated or corrected. Unfortunately, a~speckle has the same appearance as an exoplanet, and persistent speckles can easily overwhelm a faint exoplanet image. Generally, there are two approaches to dealing with speckles: control them or remove them through special data collection and processing techniques. Even better, one could use both approaches. So, we can break these techniques into three main groups, each based on the comparison of simultaneous images at different wavelengths or images of the same star at different orientations. In~particular, we discuss the Angular Differential Imaging, the~spectral differential imaging, and~the polarization~method.

\subsection{Angular Differential~Imaging}
\label{sec:adi}

\begin{figure}[H]
\includegraphics[width=10.5 cm]{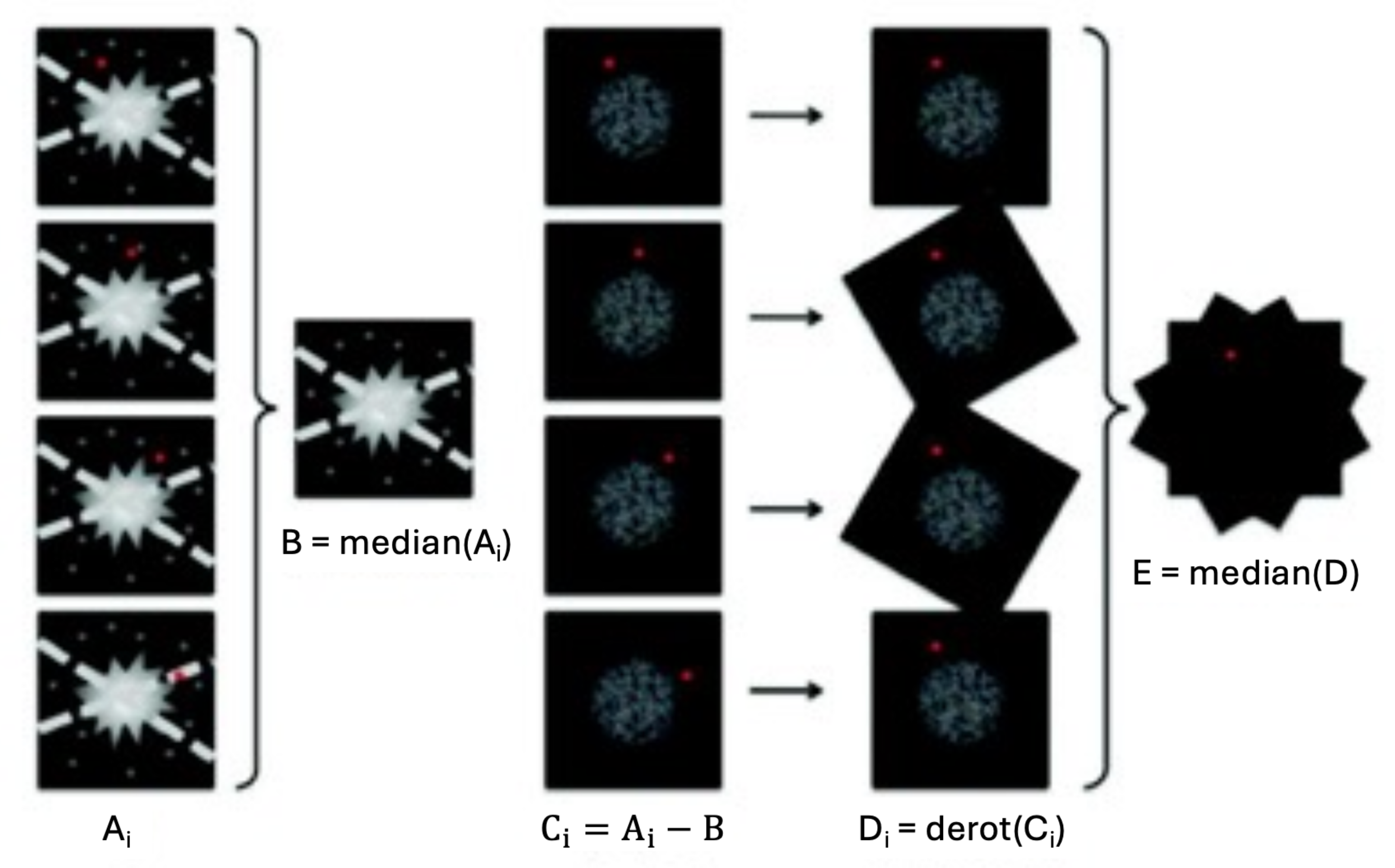}
\caption{Graphical representation of the angular differential imaging method. The red dot indicates the position of a possible planet. The~figure is taken from \url{http://web.archive.org/web/20150915005746/http://www.mpia.de/homes/thalmann/adi.htm} by Thalmann..
\label{f:adi}}
\end{figure}

The angular differential imaging (ADI) method is implemented, allowing the rotation of the field of view (FoV) centered around the target star during an observation. In~such a way each physical object in the FoV will have a different position in each frame obtained during the observing period (see the red spot in Figure~\ref{f:adi}). On~the contrary,  the~speckle pattern present in the image will not rotate during the observations. This type of observation is also said pupil stabilized as the telescope pupil does not rotate. In~principle, a median image of such a datacube should contain all the information about the speckle pattern, excluding any signal of the planets, and subtracting it from the original data should eliminate the speckle noise while retaining and highlighting the signal from any physical object (e.g., a~low-mass planet) present in the FoV at the same time. In~practice, however, the~speckle pattern is stable only for a few minutes, so a simple median of all the datasets will result in an imperfect reconstruction of the speckle pattern and, as a consequence, an imperfect subtraction of the speckle noise. When creating the image with the speckle pattern model to be subtracted from a science image we have to select just those images of the dataset near enough in time that the speckle pattern does not evolve in a substantial way. 
{A second difficulty to be overcome is that for images taken at a small angular separation from the science one (less than $\sim$$\lambda$/D), we will experience a subtraction of the signal of the low mass companion (referred to as self-subtraction)}.
The choice of the images to be used to create the speckle pattern model to be subtracted from the science image has then to be a careful trade-off between choosing images near enough in time to the science image to avoid substantial speckle pattern evolution  and with sufficient different rotation to avoid self-subtraction. This trade-off is normally different at different  separations from the host star. It is, however, important to stress that some self-subtraction of the low-mass companion signal is unavoidable, and it has to be carefully corrected to correctly evaluate the photometric signal from this object. One of the most common methods for performing this is the introduction of the original dataset of simulated companions at a known position and with known brightness. After~performing the same data reduction procedure on this  simulated dataset, it is possible to calculate the self-subtraction due to the method at different separations from the host~star.

\subsection{Spectral Differential~Imaging}
\label{sec:sdi}

\begin{figure}[H]
\includegraphics[width=13.5 cm]{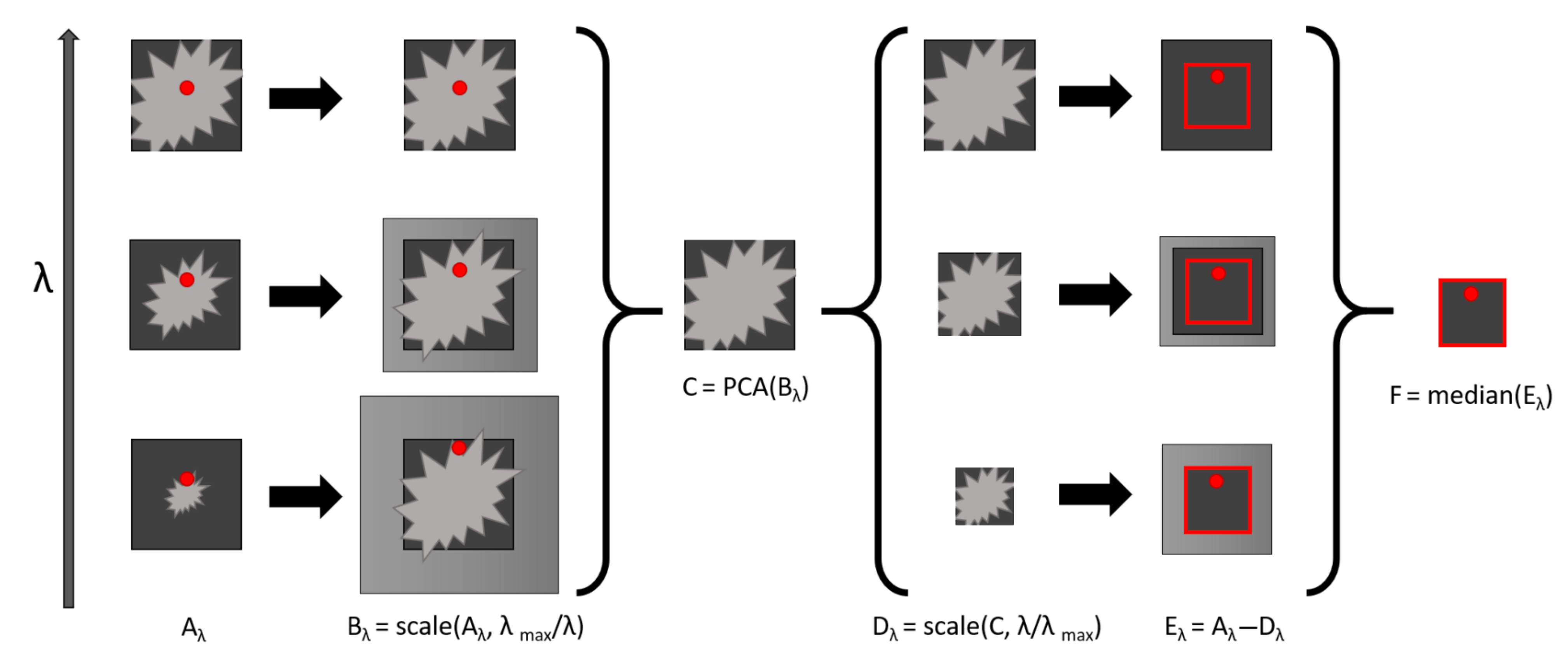}
\caption{Graphical representation of the spectral differential imaging method. The red dot indicates the position of a possible planet. Figure is taken from (Figure 1 \citet{kieferetal2021AA652_A33}).\label{f:sdi}}
\end{figure}

Spectral differential imaging (SDI) is performed when a dataset obtained simultaneously at different wavelengths is available. Indeed, as~shown in Figure~\ref{f:sdi}, the~speckle pattern scales with the wavelength, while a physical object does not change its position with the wavelength. It is then possible to rescale each image of a dataset taken at the same time to a single reference wavelength and subtract the speckle pattern. Once this is performed, each single image can be rescaled back to the original wavelength dimension and appropriately combined. As~for the case of ADI, SDI also introduces some self-subtraction that has to be appropriately calculated and corrected using methods similar to those described for~ADI.

\vspace{-6pt}

\subsection{Polarization}
\label{sec:pol}
Polarimetry is useful for observations of circumstellar disks and extra-solar planets because the light reflected from disks and planets is polarized. The~reflected light polarization from extrasolar planets is typically high, >10\% even for large separations, mainly if the phase angle is in the range 60--120$^\circ$ (e.g., \cite{seageretal2000apj540_504, stametal2004aa428_663}). On~the contrary,  no polarization is expected from the central star~\cite{schmidetla2006diesconf_165}. This situation makes dual-mode polarimetric imaging perhaps the most effective technique for speckle suppression. In~particular, the~polarization is high at wavelengths < 1 $\upmu$m due to Rayleigh scattering by atoms and molecules and scattering by aerosol particles in planetary atmospheres and dust grains in circumstellar disks. This case defines the visible as the working wavelength range for this kind of instrumentation (e.g., ZIMPOL \cite{2018A&A...619A...9S}). In~dual-mode polarimetric imaging, the~image is split by a Wollaston prism with perpendicular polarization vectors in slightly different directions, forming two simultaneous images. If~each image is subtracted from the other, only the light that is actually polarized remains in the resulting image. Because~speckles are formed by unpolarized starlight, they will be removed along with the unpolarized starlight. This technique achieved good results in imaging disks of dust that polarize light through the scattering process~\cite{kuhnetal2001apj553_L189, perrinetal2004sci303_1345, oppenheimeretal2008apj679_1574}.

\subsection{Other~Methods}
\label{sec:other}
The possibility to image young accreting planets allows us to detect these objects during their formation phase. The~most important tracer for the accretion is the hydrogen lines emission coming from the accretion shock (e.g., \citep{aoyamaetal2020arXiv201106608}). Up~to now the most used of these lines is that from the $H_\alpha$ emission as it displays the strongest signal. Detailed models of this emission were recently presented, e.g.,~by \citet{aoyamaetal2018apj866_84} and \mbox{\citet{aoyamaetal2020arXiv201106608}} resulting in a different behavior with respect to what shown by protostars (e.g., \citep{rigliacoetal2012aa548_A56}). The~main limitation of the $H_\alpha$ emission, however, is the strong expected absorption mainly due to the dust opacity that can be 10 times higher a these wavelengths than for the \mbox{NIR \citep{marleauetal2022aa657_A38}.} This absorption could explain the low number of objects detected up to now using this technique. A~possible solution to overcome this problem could be to observe hydrogen lines in the NIR where the lower absorption could compensate for the lower expected flux as in the case of the $Br\alpha$ or of the $Br\gamma$ lines in the K- and in the L-spectral bands, respectively. \par
As seen above, one of the main difficulties in the imaging of planets still embedded in the protoplanetary disk is due to the strong absorption of the signal from the planets. In~recent years, a technique allowing to overcome this problem has emerged exploiting kinematics signatures caused by planet-disk interactions (see, e.g., \citep{2022arXiv220309528P} for a review). While this is not strictly an HCI method, it allows us to define both a rough position of the companion and its mass relying mainly on the capability of ALMA to image molecular emission at high both spatial and spectral resolution. Recently, this technique led to the detection, e.g.,~of a low mass companion around HD\,142666 \citep{2023ApJ...947...60T} and RXJ1604.3-2130\,A \citep{2023A&A...670L...1S}.

\section{Instrumentation}
\label{sec:instr}
The development of the search for exoplanets through HCI has involved the preparation of an increasing number of instruments that are completely or in part dedicated to this type of observation. In~this Section, we will present some of the ground-based instruments that are at the moment in activity (see Table\ \ref{tab:instruments}), with, when they exist, their planned upgrades. We also discuss space-based instruments and  future instrumentation. 

\begin{table}[H] 
\begin{scriptsize}
\caption{High-contrast imagers on~duty. \label{tab:instr}}
\newcolumntype{C}{>{\centering\arraybackslash}X}
\newcolumntype{L}{>{\raggedright\arraybackslash}X} 

\begin{adjustwidth}{-\extralength}{0cm}
\begin{tabularx}{\fulllength}{llCCCCl}
\toprule
\textbf{Instrument}                     &\textbf{Tel.}            &\textbf{Wavel.}         &\textbf{Ang. Resol.} &     \textbf{N}      & \textbf{IWA }    & \textbf{Coronagraph}\\
                                                  &                              &   \textbf{(\boldmath{$\upmu$}m) }           &  \textbf{(mas) }                &    \boldmath{$\lambda/D$} &  \textbf{arcsec}             &                                    \\
\midrule

ERIS-NIX                                   & VLT-UT4              &   1--5                       &     25.0--129.0       &             0.5--3.0   &      0.013--0.37  &    SAM/APP              \\
GPI                                            & Gemini S.             & 0.9--2.4             & 24--62               &           4               &            0.092--0.248              &Lyot/APLC             \\
GPI2.0                                       & Gemini N.             & 0.9--2.4               & 24--62               &        3              &            0.092--0.248             &Lyot/APLC             \\
MagAO-X                                   & Magellan II           & 0.5--1.0             & $\sim$20--32        &      2--5.5    &        0.032--0.177                 &Lyot/PIAACLC/vAPP\\
MIRI                                           &  JWST                  &   10.0--23.0           &    317.0--730.0        &           1--3.3    &      0.33--2.16     &Lyot/4QPM             \\
NIRCam                                     &  JWST                  &   2.0--5.0                &    $63.0--159.0$          &            2, 4, 6     &     0.14--0.89     &Lyot                         \\
NIRISS                                       &  JWST                 &     2.8--4.8               &    88.0--152.0        &               1          &      0.12--0.15     &   AMI                         \\
NIRC2                                         & Keck-II                &     1.0--5.0 
               &   23.0--114.0        &        1.4              &    0.125                   & Lyot/VVc \\
SCExAO                                    & Subaru                 & 0.4--2.4             &10--62               &           1               & 0.010--0.062  & Lyot/PIAACMC/vAPP/vortex   \\
SHARK-VIS                               &  LBT                     &   0.4--0.9               & 11--20                    &        2.5--10        & 0.050--0.200                         &      Lyot/Gauss                            \\
SHARK-NIR                               &  LBT                     &   0.96--1.7              &24--42                   &                           &     0.050--0.150            & Lyot/SP/4QPM              \\
SPHERE                                    &  VLT -UT3            & 0.95--2.32        & 24--62               &       1, 2, 3         &      0.072--0.180 &  Lyot/APLC/4QPM \\

\bottomrule
\end{tabularx}
\end{adjustwidth}
\noindent{\footnotesize{4QPM: 4 Quadrants Phase Mask; AMI: Aperture Masking Interferometry; APLC: Apodized Pupil Lyot Coronagraph; APP: Apodizing Phase Plate; PIAACLC: Phase Induced Amplitude Apodization Classic Lyot Coronagraph; PIAACMC: Phase Induced Amplitude Apodization Complex Mask Coronagraph; SAM: Sparse Aperture Mask; SP: Shaped pupil;  vAPP: vector Apodizing Phase Plate; VVc: Vector Vortex Coronagraphic mask.}}
\label{tab:instruments}
\end{scriptsize}
\end{table}
\unskip

\subsection{Present~Instrumentation}
\subsubsection{GPI and~GPI2.0}
The Gemini planet imager (GPI; \citep{macintoshetal2014pnas11112661}) was installed at the Gemini South telescope (\mbox{8.2 m}) between 2014 and 2020. It operated in the NIR with a wavelength range between 0.9 and 2.4~$\upmu$m.
 It had an integral Field Spectrograph (IFS) with R$\sim$40, and~it also provided the possibility to perform dual-channel polarimetry on the full field in Y, J, H, and~K spectral bands.  Also, this instrument is equipped with an ExAO with a DM of 50 actuators per beam diameter and a Shack-Hartmann wavefront sensor ($1 \leq \text{speed} \leq 2$\ kHz) as primary WFS. 
GPI, decommissioned in August 2020,  is currently in the process of being upgraded to a new version named GPI\,2.0, and~the first light is expected at the Gemini---North telescope in late 2024 or early 2025 \citep{chilcoteetal2024ess562807}. This new version will have new coronagraphs with smaller inner working angles, a~new wavefront sensor at higher sensitivity, and~the possibility to use new spectroscopic observing modes. The~goal of the new instrument is to obtain contrasts of the order of $10^{-7}$ at separations lower than 1" from the target~star.

\subsubsection{SPHERE and SPHERE+}
SPHERE (Spectro-Polarimetic High-contrast imager for Exoplanets REsearch, \citep{beuzitetal2019aa631_A155}) is a planet imager installed at ESO/VLT (8.2 m) since 2015. It is composed of three different scientific modules. The~first one is the IFS  \citep{claudi2008SPIE.7014E..3EC} that operates between the Y and the J band (0.95--1.35~$\upmu$m) with a spectral resolution of R$\sim$50 or R$\sim$30 in the Y and the H spectral bands (0.95--1.65~$\upmu$m), according to the observing mode. 
The second instrument is the Infra-Red Dual-beam Imager and Spectrograph (IRDIS; \citep{dohlen2008SPIE.7014E..3LD}) that can operate both together with the IFS or as a stand-alone instrument. IRDIS has both narrow- and broad-band spectral filters ranging between the J and the K spectral bands.
IRDIS can operate as a dual-band imager adopting two narrow-band filters at adjacent wavelengths \citep{vigan2010MNRAS.407...71V}, as~a classical imager on a single spectral filter, and~also as a polarimetric instrument (see, \mbox{e.g., \citep{2020A&A...633A..63D, 2020A&A...633A..64V}}). Finally, long-slit spectroscopy (LSS; \citep{2008A&A...489.1345V}) allows one to obtain medium resolution (R$\sim$350) spectra of bright  companions at a separation of a few tenths of an arcsec from the host star. The~third instrument is ZIMPOL (Zurich Imaging Polarimeter; \citep{2018A&A...619A...9S}), a polarimeter operating in the visible between 500 and 900~nm. 
SPHERE is also equipped with an ExAO system called SAXO (SPHERE extreme AO system) with a DM of 50 actuators per beam diameter and a Shack-Hartmann WFS ($1 \leq \text{speed} \leq 2$\ kHz) as primary WFS \citep{2006OExpr..14.7515F, 2014SPIE.9148E..1UF}. In~the next years, SPHERE will upgrade to SPHERE+ \citep{boccalettietal2022spie12184E_1S} that consists of a new XAO system stage named SAXO+. The~new AO stage will operate in the NIR and at a higher frequency (2--3~kHz) than the SAXO system. It has been designed to operate with both IRDIS and IFS but it will be able to do the same also for other future instrumentation. The~second element of the upgrade should be a medium-resolution IFS named MEDRES \citep{2022SPIE12184E..4FG}, which should provide medium-resolution (R$\sim$1000) spectra on a limited FoV ($\sim$0.4$"$ squared) and a wavelength ranging between 1.2 and 1.65~$\upmu$m. This will allow us to reach a contrast of $10^{-5}$--$10^{-6}$ down to separations of 0.1$"$--0.2$"$. However, this second element is in stand-by waiting for a future~approval.

\subsection{NIRC2 and~KPIC}
NIRC2 is a near-infrared instrument operating since Summer 2001on the Keck II telescope (10.0 m). The~instrument \endnote{\url{https://www2.keck.hawaii.edu/inst/nirc2/tech_home.html}, \hl{accessed on 31 August 2024.}} operates from 1 to 5 $\upmu$m, with~three observing modes (OM). Imaging OM exploits three different cameras that provide scales of 0.01, 0.02, and~0.04 arcsec/pixel close to the Keck diffraction size at J, K, and~M bands. The~coronagraphic OM is characterized by the presence of several Lyot masks and two vector vortex coronagraphic (VVC) masks, one optimized for the  L-/M-band, and~the second for the K-band. Finally, the~third OM is the grism spectroscopy with several low and medium dispersion grisms. The~NIRC2 underwent the two phases upgrade named Keck Planet Imager and Characterizer (KPIC; \citep{wangetal2024arXiv240615028W}) with the installation of an infrared pyramid WFS (PyWFS) based on a fast, low-noise SAPHIRA IR-APD array, and~an array of single-mode fibers with the aid of an active fiber injection unit (FIU) which feed the NIRSPEC spectrograph with \mbox{R >30,000} (first Phase). The~second phase of the upgrade consisted of the installation of a 1000-actuator deformable mirror, beam-shaping optics, a~vortex coronagraph, and~other upgrades to the FIU/FEU \citep{wangetal2024arXiv240615028W}.

\subsubsection{ERIS}
ERIS (Enhanced Resolution Imaging Spectrograph; \citep{2023A&A...674A.207D}) is an imager and a spectrograph recently installed at VLT (the first light was in February 2022). It was conceived to replace two ESO instruments SINFONI and NACO. One of the science cases of ERIS is the  detection of very planets and the characterization of the giant gaseous planet \mbox{atmospheres \citep{2023A&A...674A.207D}.} ERIS is composed of two science sub-systems. The~first one, called \mbox{NIX \citep{pearsonetal2016SPIE9908E_3},} is an imager operating between 1 and 5~$\upmu$m. NIX  operates both in coronagraphic and not coronagraphic mode allowing the  choice between a large selection of both broad- and narrow-band filters. Finally, it has an LSS mode with R$\sim$450. The~second scientific subsystem is SPIFFER, which operates at wavelengths between 1 and 2.5~$\upmu$m
at spectral resolution R$\sim$5000 over the full band or R$\sim$10,000 over the half band.  Furthermore, an~updated AO system based on the UT4 Adaptive Optics facility (AOF, \citep{riccardietal2022SPIE1117_122629425}) feeds~ERIS.

\subsubsection{SCExAO}
Subaru Coronagraphic Extreme Adaptive optics (SCExAO; \citep{2009SPIE.7440E..0OM}) is installed at the Subaru Telescope (8.3 m) with the main goal of imaging exoplanets and disks around nearby stars. The~light is first processed by the AO188 adaptive optics system to remove most of the optical aberrations induced by atmospheric turbulence. The~optical beam is then processed by SCExAO (with a DM of 48 actuators per beam diameter \citep{jovanovichetal2015PASP127_890}),  which performs fine wavefront correction (ExAO) and removes the bright  starlight exploiting a suite of high efficiency and low inner working angle (IWA) coronagraphs. The~light from SCExAO is then provided to different scientific modules, amidst which the Coronagraphic High Angular Resolution Imaging Spectrograph (CHARIS; \citep{2015SPIE.9605E..1CG}) is an IFS covering a wavelength range in the J, H, and~K spectral bands. Its spectral resolution is R$\sim$20 using all three spectral bands and~between 70 and 90 for each singular spectral band alone. CHARIS can also operate in the spectro-polarimetric observing mode. SCExAO can feed other instruments like the Visible Aperture Masking Polarimetric Imager for  Resolved Exoplanetary Structures (VAMPIRES \citep{2014SPIE.9146E..0UN}) that operates in the visible spectrum at wavelengths between 600 and 800~nm. 

\subsubsection{SHARK}
SHARK (System for coronagraphy with High order Adaptive optics from R to K band) is one of the Large Binocular Telescope (LBT, $2\times 8$\ m) instruments and~consists of a pair of instruments working in visible  (0.4--0.9 $\upmu$m, SHARK-VIS ) and in the near-infrared (\mbox{0.96--1.7 $\upmu$m,} SHARK-NIR), with~the possibility of working in~parallel. 

The main innovation of SHARK-VIS \citep{mattioli2018SPIE10702E..4FM}, installed at LBT in 2023, is the implementation of a high-cadence-imaging observing mode ranging from 1~KHz to 65~Hz. This observing mode has been tested at the telescope with a forerunner instrument  \citep{pedichini2017AJ....154...74P,licausi2018SPIE10703E..2UL} that demonstrates the capability of the method to obtain a contrast of the order of $10^{-5}$ down to separations of the order of 100~mas around bright stars. 

SHARK-NIR \citep{farinatoetal2022spie12185E_22} is a high-contrast imager installed on the left arm of the LBT. SHARK-NIR takes full advantage of the extreme AO correction delivered by SOUL (Single Conjugated Adaptive Optics Upgrade for LBT; \citep{pinna2021arXiv210107091P}) that has recently been installed on the  LBT. This new AO system is based on a Pyramid WFS allowing to obtain a SR$\sim$70\% for as faint stars as R$\sim$13~mag. SHARK-NIR will operate in the Y, J, and~H spectral bands.
A suite of different coronagraphs is available for different types of scientific  cases. The~instrument is also equipped with both wide and narrow band filters with the possibility to perform dual-band imaging. Finally, the~instrument performs low-resolution (R$\sim$100) and medium-resolution (R$\sim$700) spectroscopy using the LSS mode with the possibility of spectrally characterizing the atmospheres of bright substellar companions with a contrast of the order of $\sim10^{-5}$ at separations larger than a few tenths of~arcsec.   

\subsubsection{MagAO-X}
MagAO-X is the extreme adaptive optics system for the 6.3 m Magellan II (Clay) with a high-order control loop $> 2$ kHz consisting of a 97 actuator woofer deformable mirror (DM), a~2040 actuator tweeter DM, and~a modulated pyramid wavefront sensor \mbox{(WFS) \citep{malesetal2020SPIE11448E_4L, close2018spie10703E_4Y}.}  MagAO-X is equipped with several coronagraphs and~two science cameras,  so it is  possible to carry out science observations in two filters simultaneously (dual-band imaging).  The~main science goal of MagAO-X is the search for and the characterization of young accreting planets in H$_\alpha$ orbiting nearby T Tauri and Herbig Ae/Be stars \citep{close2020aj160_221}.  

\subsubsection{Medium and High Resolution~IFSs}
On several 8 m class telescopes are mounted NIR IFS instruments with medium and high resolution. Some of them are (the list is not complete): MUSE@VLT, OSIRIS@Keck, NIFs@Gemini. These instruments have been used to characterize spectroscopically, among others, early L-type companions at an angular separation $>0.5"$ (\citep{currieetal2023ASPC534_799} and reference therein).

\subsection{Space Based~HCI}
Even though it was not built for HCI observations, the~Hubble Space Telescope (HST) has proven to be a tool for HCI due to its relatively stable point-spread-function (PSF) which allows for the subtraction of light from the central source using a variety of \mbox{methods \citep{krist2004spie54871284K}.} In~2005, \citet{lowranceetal2005aj130_1845} (see Section
 \ref{sec:statistics}) planned to use a Near-Infrared Camera and Multi-Object Spectrometer (NICMOS) to discover substellar companions to young, nearby stars. NICMOS is a second-generation instrument for the HST and has three available cameras. Camera 2 was equipped with a coronagraph, which is a hole in the field divider mirror that is baffled with a cold pupil plane mask. The~NICMOS was no longer offered after cycle18 (2010) for observations. Nevertheless, people also continued to use HST instrumentation for HCI observations in the last few years; for~example, \citet{cugnoetal2023aj166_162} used data from Mag AO-X together with HST/WFC3 data to observe the emission of H$_\alpha$ from a candidate circumplantary disk orbiting the young K5 star V 1121 Oph. The~Wide Field Camera 3 (WFC3) has been in operation since 2009. It has two channels; one UVIS operates from about 200 to 1000 nm. The~second is an IR channel that operates from 900 to 1700~nm. 

At the end of 2021, the James Webb Space Telescope (JWST) was launched toward a solar orbit near the Sun-Earth L2 Lagrange point, where it finished its commissioning phase in the summer of 2022.  The~JWST  has a focal plane assembly composed of four scientific modules operating in NIR and MIR wavelength range. Three of them have OMs configured also for HCI observations. Two of them operate in the NIR:  the NIRCam coronagraphic imaging OM \citep{2010SPIE.7731E..3JK} with five different Lyot-type coronagraphs allowing IWA between 0.14$"$ and 0.89$"$, and~NIRISS Aperture masking interferometry  \citep{sivaramakrishnanetal2023pasp135_5003} working in the wavelength range between 2.8 and 4.8~$\upmu$m.This latter mode turns the full aperture of JWST into an interferometric array as the light is admitted by seven sub-apertures to produce an interferogram on the detector. This mode allows to detect and characterize planetary or stellar companions that are up to ~9 magnitudes fainter than their host star and separated by \mbox{$\sim$70-- 400 mas.} The~MIRI coronagraphic imaging mode \citep{2015PASP..127..633B} operates in the MIR exploiting 4QPM coronographs with IWA of  0.33$"$, 0.36$"$, 0.49$"$, and 2.16$"$ for a Lyot~mask.

\subsection{Future HCI~Instrumentation}
In the next years, in~addition to the several upgrades of existing instrumentations discussed in the previous section, several brand new high-contrast imagers will be mounted on 8--10 m  and~on the extremely large class ($\sim$40 m) telescopes. For~the former, it worths to mention the IFS FRIDA (inFRared Imager and Dissector for Adaptive optics) at the Grantecan telescope (10 m) that will be in operation at mid 2025\endnote{\url{https://www.gtc.iac.es/instruments/instrumentation.php}, \hl{accessed on 30 September 2024.}}. FRIDA will work in the near-infrared wavelength ranges (0.9--2.5 $\upmu$m) with imaging capability. It will make use of the GTC Adaptive Optics system (GTCAO). The~IFS spectral resolution, depending on the wavelength, ranges between 1000 and 32,000. 

The extremely large telescopes, both the TMT (Thirty Meter Telescope)\endnote{\url{https://www.tmt.org/page/second-generation-instruments}, acccessed on 30 September 2024.}, and~the E-ELT (European Extremely large Telescope)\endnote{\url{https://elt.eso.org/instrument/} \hl{acccessed on 30 September 2024}.} are foreseen to have high-contrast imagers. For~the TMT, the~second generation set of instrumentation is foreseen for two planet imagers. The~first is a mid-IR spectrometer imager IFU named bMICHI (Mid-IR Camera, High-disperser and IFU spectrograph), that will operate an IFS mode in the  3--14 $\upmu$m with R$\sim$1000, and~a long slit and polarimetric mode with resolution ranging between R$\sim$600 and \mbox{120,000  \citep{packhametal2018spie10702_EA0}.} The~second is the Planetary System Instrument (PSI; \citep{jensenclemetal2021SPIE11823_E09}), which is a high-contrast instrument that will work between 0.6 and 14 $\upmu$m (three arms PSI-Blue: 0.6--1.8 $\upmu$m, PSI-Red: 2.0--5.1 $\upmu$m, and~PSI-10: 7.0--14 $\upmu$m) with a resolution from R$\sim$50 to R$\sim$6700 .  PSI will be equipped with several coronagraphs (vAPP, Vector-Vortex $+$ Lyot Stop).

The first generation instrumentation of the E-ELT will comprise two planet imagers and spectrographs: METIS (Mid Infrared ELT Imager and Spectrograph), and~HARMONI (High Angular Resolution Monolithic Optical and Near---Infrared Integral field spechtrograph). METIS will work with a set of four different focal planes grouped in three different subsystems: the imager (IMG) and the spectrograph (LMS) are the two scientific focal planes, and~the last one, SCAO, is the dedicated adaptive optics system \citep{serraetal2024SPIE13103E1US}. METIS will be encased in a cryostat and will work between 3 $\upmu$m and 13.5 $\upmu$m with its imaging arm, while the spectroscopic arm, an~IFS with R$\sim$100,000, will work between 3 $\upmu$m and 5 $\upmu$m, assisted by a coronagraph. HARMONI will cover \citep{thatteetal2024SPIE13096E14} a spectral range from 0.450 $\upmu$m to 2.450 $\upmu$m with R from 3500 to 18,000 and spatial sampling from 60 mas to 4 mas. It can operate in two Adaptive Optics modes---SCAO (including a high-contrast capability) and LTAO---or with noAO (no use of AO).  Further in time, the~second generation of E-ELT instrumentation foresee the Planetary Camera and Spectrograph (PCS), a~concept that is the combination of eXtreme Adaptive Optics (XAO), coronagraphy and spectroscopy \citep{kasperetal2021Msngr_182_38} to detecting and characterizing nearby exoplanets with sizes from sub-Neptune to Earth-size in the neighborhood of the~Sun. 

Scheduled to launch in the mid-2020s, the~Nancy Grace Roman Space Telescope\endnote{\url{https://science.nasa.gov/mission/roman-space-telescope/introducing-the-roman-space-telescope/}, \hl{accessed on 30 September 2024.}} will mount an HCI. It is a NASA observatory with a primary mirror of 2.4 m. It will have two instruments: the wide field instrument (WFI), and~a coronagraphic instrument technology demonstration which will perform HCI and spectroscopy for the discovering and charaterization of extrasolar planets \citep{spergeletal2013arxiv13055425}.

\begin{figure}[H]
\includegraphics[width=10.5 cm]{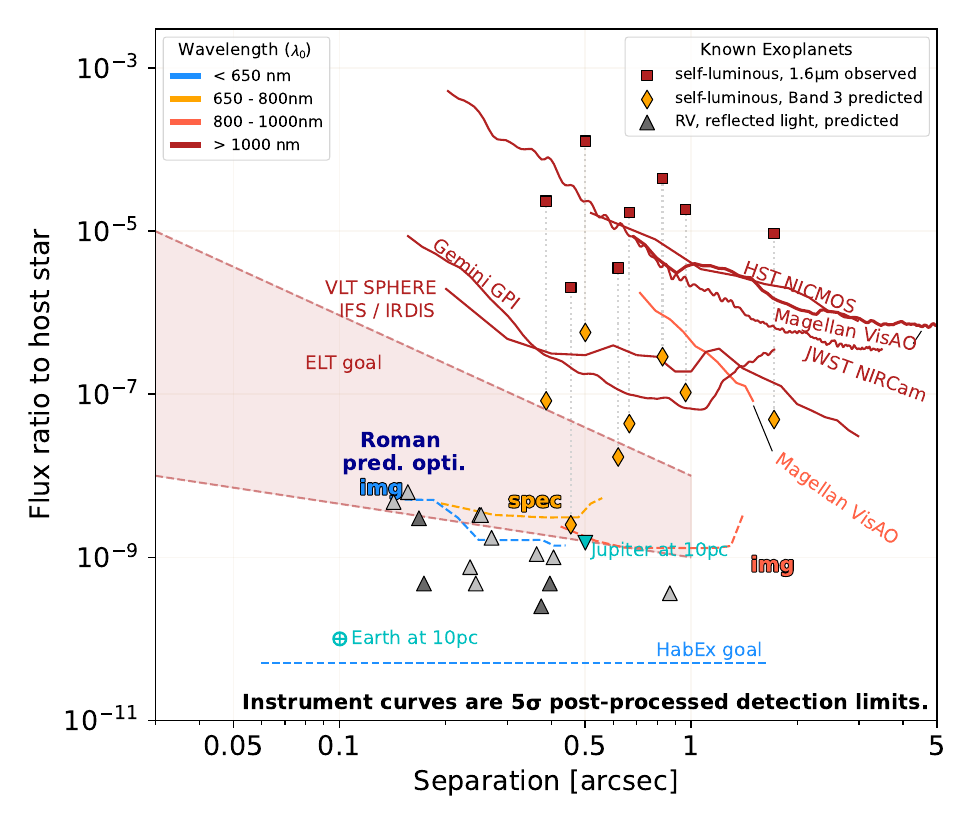}
\caption{The $5\ \sigma$ post-processed contrast curves of several both ground- and space-based high-contrast imagers. Code and data source by V. Bailey and S. Hildebrandt Rafels (\url{https://github.com/nasavbailey/DI-flux-ratio-plot}, accessed on 30 September 2024)\label{fig:instrcontrast}}
\end{figure}

Figure\ \ref{fig:instrcontrast} compares the measured and predicted performances of most of the instruments described previously in~several wavelength regimes. Moreover, both detected (red squares) and simulated (orange diamonds) planets are also over-plotted to allow us to understand the limits of these instruments in planet detection. The~future ELT and new space-based instruments will allow us to reach the contrast value necessary to observe planets in reflected light (grey triangles), even at short angular~separations.

\unskip 

\section{Algorithms}
\label{sec:sdec}
A number of different algorithms have been created in order to implement the high-contrast imaging methods described above.
Here, we will give a short description of some of them with the advice that, given the huge development of this field in
recent times, the~list could be less than~complete. \par
One of the first algorithms adopted for the subtraction of the speckle noise and one of the most widely used in the past years
for HCI is the locally optimized combination of images (LOCI; \citep{2007ApJ...660..770L}). The~aim of the method is to
construct an optimized reference image to be subtracted from the science image, maximizing the speckle subtraction and 
minimizing the self-subtraction of possible companions. This is obtained by dividing the image into subsections and creating, 
independently for each subsection, a~linear combination of reference images in such a way that their subtraction from the 
science image could minimize the speckle noise. A~correct optimization of the weights of each reference image can, moreover, 
help in reducing the self-subtraction of any science target in the FoV. While LOCI is particularly well optimized to reduce
data using the ADI method, it can create large self-subtraction when also spectral information is introduced using SDI 
data. To that end, an update of this algorithm has been introduced with TLOCI (template LOCI; \citep{2014IAUS..299...48M}).
This algorithm takes into account the possible large flux variation at different wavelengths present in a planetary 
spectrum. Moreover, it couples to the original LOCI algorithm with the possibility of maximizing the SNR of a companion with
a specific spectrum adopted as a template by the algorithm itself. A~further update of the algorithm was proposed with
MLOCI (matched LOCI; \citep{2015A&A...581A..24W}) that included in the science data simulated point source used to 
maximize the detection~SNR. \par
An alternative and widely used algorithm is based on the principal components analysis (PCA) of the Karhunen–Loeve (KL)
transformation of a library of reference PSF images. It was described for the first time in \citet{2012ApJ...755L..28S}
together with the Karhunen--Loeve  image projection (KLIP) algorithm. After~selecting them in an appropriate way for the PSF library,
the algorithm calculates their KL transform, selects the number of modes to be retained, and~uses them to reconstruct the
PSF image that will be subtracted from the science image. An~alternative method to implement the PCA is to use the 
single value decomposition (SVD) to generate three arrays, which, if combined correctly, lead to the eigenvectors and 
the eigenvalues used to reconstruct the original data. A~first implementation of this method has been used for the 
Python-based tool called PYINPOINT \citep{2012MNRAS.427..948A}. The~same method was also used in 
\citet{2015A&A...576A.121M} for the reduction of the data of the IFS of SPHERE and later included in the SPHERE data 
reduction pipeline called SpeCal \citep{2018A&A...615A..92G}. Independent of the technique used to implement it, PCA
needs a careful choice of the number of principal components to be used to create the reference image that will be
subtracted from the science data. Indeed, when using a large number of principal components, the~speckle pattern will 
be reconstructed with very good precision, allowing a good subtraction of the speckle pattern, but, at~the same time, it will
result in a strong self-subtraction. On~the contrary, using a too-low number of principal components will result in a bad
speckle subtraction, impeding the detection of point sources with acceptable SNR. A~trade-off in the choice of the number
of principal components is then fundamental, and it can depend on several different factors, e.g.,~the separation from
the central star and if we want to image a point source or an extended structure (e.g., a~protoplanetary disk). A~possible
solution to such a problem is the introduction of an annular PCA procedure that performs different PCA at different separations
from the host star, as performed in the VIP (Vortex imaging \mbox{processing; \citep{2017AJ....154....7G}}) pipeline.  \par
Different alternative algorithms have been proposed in the past few years with the aim to enhance the SNR of the detection
reducing (or nulling) at the same time the self-subtraction linked to the ADI method. One such algorithm is the PAtch
COvariance (PACO; \citep{2018A&A...618A.138F}) method that is based on the definition of a nonstationary multi-variate
Gaussian model of the background directly from the data. Once this is performed, a~binary hypothesis test is applied to decide
on the presence of a planet. An~update of the same algorithm (PACO ASDI; \citep{2020A&A...637A...9F}) also allowed its 
application to collect data with spectral information like those from an IFS. More recently, the algorithm was also applied to
the detection of extended structures like circumstellar disks (REXPACO; \citep{2021A&A...651A..62F}). Finally, very 
recently the same authors developed a further improvement of the method coupling the original PACO algorithm with supervised
deep learning (deep PACO; \citep{2023arXiv230302461F}) with the aim to reduce the uncertainties linked to the approximate 
fidelity of the PACO statistical model to the time evolving~observations. \par
The detection of substellar companions in a dataset previously reduced using the ADI technique is an objective that can be
achieved by exploiting apposite algorithms like ANDROMEDA (ANgular Differential OptiMal Exoplanet Detection Algorithm; \citep
{2015A&A...582A..89C}). This algorithm searches for the signature of a companion in ADI-reduced differential images by
applying a maximum likelihood estimation of its position and intensity. A~similar approach was followed for the reduction 
of the GPI data previously reduced using the KLIP algorithm by \citet{2017ApJ...842...14R}. They demonstrated that the application of a forward model matched filter (FMMF) allowed for a most effective recovery of a planetary signal while reducing the false-alarm~probability at the same time. \par

A similar approach is followed by the regime-switching model (RSM; \citep{2020A&A...633A..95D,2021A&A...646A..49D}) 
detection map. It is based on a model developed more than thirty years ago in the field of econometrics, and it allows one
to use the residual cubes obtained by applying different ADI-based techniques to create a single detection map. The~algorithm
was then further improved by the same authors by allowing an automated optimal selection of the parameters to be used 
in the algorithm itself (Auto-RSM; \citep{2021A&A...656A..54D}). The~creation of a reliable detection map is also the main 
objective of the standardized trajectory intensity mean map (STIM \mbox{map \citep{2019MNRAS.487.2262P}}) algorithm. It is 
different from the detection map normally found in literature as they assume that the residual noise in an ADI-reduced image
is Gaussian. As~the authors point out, this can lead to a large number of false-positive detections. To~solve this
problem they assume for the speckle noise a Modified Rician distribution, which allows for a strong reduction of false-positives. \par
Recently, some algorithms were developed to solve some specific problems of the HCI. One of these is the
Morphological Analysis Yielding separated Objects in Near infrAred usIng Sources Estimation (MAYONNAISE; \citep
{2021MNRAS.503.3724P}) algorithm. It is mainly devoted to the imaging of circumstellar disks, avoiding the distortion normally 
induced by the ADI technique and preserving their shapes and flux distribution. Another one is the Temporal Reference
Analysis of Planets (TRAP; \citep{2021A&A...646A..24S}), which is mainly aimed at detecting planetary mass companions at short
(<3$\lambda/D$) separation from the host~star.

\section{Results}
\label{sec:res}
In 1995, \citet{nakajimaetal1995natur378_463} announced the discovery of a sub-stellar object around a nearby M1V star GJ\,229 obtained with the Adaptive Optics Coronagraph (AOC) and the Palomar 60-inch telescope. GJ\ 299\ B is a brown dwarf with a mass of 20--50~$M_{Jup}$, and~a spectral type of T7, and~a very low $T_{eff}$ of 900~K. It worths to be mentioned because, also if it is not a planetary mass companion, it was the first  object discovered with a high-contrast instrumental set-up\endnote{During the writing of this review, \citet{xuanetal2024arXiv241011953X} observing GJ\ 229 with GRAVITY interferometry resolved Gliese 229 B into two components, Gliese 229 Ba and Bb.}. After~quite a decade, the~HCI technique discovered the first planetary-mass object orbiting the brown dwarf 2M 1207 (short for 2MASSWJ\,1207334-393254) in the TW\,Hydrae association \citep{chauvinetal2004aa425_l29}. The~companion, at~an angular separation of $\sim$0.78$"$, had an estimated mass of around 5\ $M_{Jup}$ and a spectral type between L5 and L9. Since this first discovery, over~30 planets and several BD companions have been discovered with the HCI technique. Most of these planets are in unexpectedly wide orbits with hundreds and thousands of au. These wide companions resulted in a critical review of the canonical theories of planet formation via disk instability and core accretion and introduced the possibility that it represents the tail of the brown dwarf formation \citep{bowler2016pasp128_j2001, currieetal2023ASPC534_799} as opacity-limited fragments of turbolent, which collapse molecular clouds \citep{lowandlyndenbell1976MNRAS176_367, silk1977apj214_152, boss2001apj551_L167, bate2009mnras392_590}. Planets are identified as bodies with masses below the deuterium burning mass threshold (\mbox{13 M$_J$}), as stated by the IAU Working Group Definition. Nevertheless, the~recent low mass companion discoveries  show that this limit could not sharply define a boundary between planets and sub-stellar objects but~rather that it should be incorporated with the tail-end of the stellar formation. As~matter of fact, there are objects with masses below the deuterium burning threshold found in systems with an architecture that instead suggest the cloud-fragmentation formation, and~viceversa (e.g., \citep{todorovetal2010apj714l_84, liuetal2013ApJL777_L20, quirrenbachetal2019aa624_A18}). Another possible scenario is that a planet forms in a circumstellar disk around a young star, and~its mass could be proportioned to the primary mass \citep{gretherandlineweaver2006apj640_1051}. Furthermore, the~distribution of the protopalenetary disk radii has a peak at around 200 au and suddenly decreases at around 300 au. Both those situations draw some authors to define the following limits for the planet---BDs boundary: \mbox{mass < 25 $\text{M}_J$,} $\text{a}_\text{P} \leq 300$\ au, and~the ratio between the companion and the primary mass q < 0.025 \citep{currieetal2023ASPC534_799}. In~any case, the~matter is already under~debate. 

The connection between young planets and the protoplanetary disks is very important, and~in particular, several efforts have been devoted to the observations of young stars with transitional disks with cavities depleted of dust (e.g., \citep{dodsonrobinsonandsalyk2011apj738_131, espaillatetal2014prplconf_497, owen2016pasa33_5}). This approach led to the  discovery of several companions of the gaps, from~the stellar to the planetary mass regime (two examples for all: HD142527 \citep{billeretal2012apj753_L38, closeetal2014apj781_L30, rodigasetal2014apj791_L37, lacouretal2016aa590_A90, claudietal2019aa622_a96}, and~HD100546 \citep{quanzetal2015apj807_64, currieetal2015apj814_L27, garufietal2016aa588_A8, sissaetal2018aa619_A160}).

Due to the complexity of this technique, the~systems with confirmed HCI companions are fewer than the other indirect methods. Also, the surveys conducted with the last-generation instruments resulted in the detection of a low frequency of giant planets at separations larger than 10~au from the host star \citep{nielsenetal2019aj158_13,viganetal2021aa651_a72} that are the main targets for such instruments. As~a result, blind surveys such as the SHINE survey performed with SPHERE \citep{2017sf2a.conf..331C, desideraetal2021aa651_a70, langloisetal2021aa651_a71} need to observe hundreds of stars to be able to detect a handful of companion. Coupling HCI with other techniques can help in increasing the probability of detecting a sub-stellar companion. An~example of this type is the survey undertaken in the past years to detect companions discovered through RV data. Despite the low overlapping of these two methods (RV prefers older objects at small separation from their host star), this technique allowed for the detection of objects in the past few years, including, e.g.,~$\beta$\,Pic\,c \citep{nowaketal2020aa642_L2} and HD\,206893\,c \citep{hinkleyetal2023aa671_L5}. {In the last years, some authors demonstrated that coupling HCI with astrometric data deriving from the comparison of different catalogs like those from Hipparcos and Gaia to obtain the proper motion anomaly could be a new technique to discover planets} (PMa; see, e.g., \citep{2021ApJS..254...42B, 2022A&A...657A...7K}).
Furthermore, coupling these data allows us to determine the companion dynamical mass together with important orbital parameters while using a low number of DI epochs. This technique allowed for the detection of new sub-stellar companions in the past few years, including those detected in the BD regime by the COPAINS survey \citep{2022MNRAS.513.5588B} down to objects just above the deuterium~limit. 

In Table\ \ref{tab:remarkableB} are listed the more remarkable (at the flavor of the authors) objects found with HCI and DI. In~the following a short description of each object is given. Some complete lists of these objects are in the already cited review of \citet{bowler2016pasp128_j2001} and \citet{currieetal2023ASPC534_799}.

\begin{table}[H]

\caption{Some remarkable planetary systems discovered or characterized with DI and HCI. The~discovery year is indicated in the last~column}
\tablesize{\footnotesize}{
\begin{adjustwidth}{-\extralength}{0cm}
\begin{tabularx}{\fulllength}{llll}
\toprule
\textbf{Name}              &     \textbf{Description }               &  \textbf{Comments}       & \textbf{Discovery} \\
\midrule
GJ 229 B $^1$            &   1 Brown Dwarf          &1st BD discovered with HCI method &1995\\
2M 1207 $^2$        &   1 Planet                     &1st PMC found around a BD &2004\\
GQ Lup $^3$         &   1 Planet  or BD         &1st PMC found around a young star &2005\\
AB Pic $^4$            &   1 Planet                     & PMC at planet/ BD boundary &2005\\
HR8799 $^{5,6} $     & 4 Planets                     & 1st Planetary system identified with HCI and benchmark& 2008--2010\\
                                &                                     &in understanding young planetary atmospheres &   \\
$\beta$ Pic $^{7,8} $          &   2 Planets                   & 1st bona fide planet discovered in NIR and MIR in Thermal Emission &2008\\
Ross 458 (AB) $^{9,10,11}  $   &  1 Planet                     & 1st circumbinary planet imaged &2010\\
LkCa 15 $^{12,13,14,15}$     &  2 Protoplanets $^\text{a}$     &  1st protoplanet seen in process of active accretion &2011\\
51 Eri $^{16}$               &   1 Planet                    & 1st planet discovered with XAO system (GPI) &2015\\
2MJ2126 $^{17,18} $      &  1 Planet                     & The widest orbit planetary-mass object known &2016\\
HIP 65426 $^{19,20}$              & 1 Planet                      & 1st Planetary system observed with JWST (ERSP) &2017\\
PDS 70 $^{21,22,23}$                 & 2 protoplanets             & 1st detection of jovian protoplanets &2018--2019\\
HIP 99770 $^{24}$                      & 1 Planet                   & 1st Planet found with Astrometry and HCI synergy&  2023\\
AF Lep $^{25, 26, 27}$               &  1 Planet                   & 1st Planet  under the D-burning limit and a < 10 au&  2023\\
                                                     &                                 &  found with Astrometry and HCI synergy                               &         \\
$\epsilon$ Ind  $^{28}$                             & 1 Planet                   &  1st planet discovered with JWST                               & 2024 \\                                                   
\bottomrule
\end{tabularx}
\end{adjustwidth}}
\noindent{\footnotesize{$^\text{a}$ the number of protoplanets in this system is controversial, see \citet{currieetal2019apj877_L3}.
$^1$ \citet{nakajimaetal1995natur378_463}; \mbox{$^2$ \citet{chauvinetal2004aa425_l29};} $^3$ \citet{mugrauerandnehuauser2005AN326_701}; $^4$ \citet{chauvinetal2005aa438_L29}; $^5$ \citet{maroisetal2008sci322_1348}; \mbox{$^6$ \citet{maroisetal2010Natur468_1080};} $^7$ \citet{lagrangeetal2009aa493_L21}; $^8$ \citet{lagrangeetal2019NatAs3_1135}; $^{9}$ \citet{Dupuyandkraus2013sci341_1492}; $^{10}$ \citet{goldmanetal2010mnras405_1140}; $^{11}$  \citet{scholzetal2010aa515_A92};  $^{12}$ \citet{krausandireland2012apj745_5}; $^{13}$ \citet{sallumetal2015Natur527_342}; $^{14}$ \citet{bowler2016pasp128_j2001}; $^{15}$ \citet{currieetal2019apj877_L3}; $^{16}$ \citet{macintoshetal2015sci350_64};  $^{17}$ \citet{fahertyetal2013aj145_2}; $^{18}$ \citet{deaconetal2016mnras457_3191}; $^{19}$ \citet{chauvinetal2017aa605_9}; $^{20}$ \citet{carteretal2023ApJ951_L20}; $^{21}$ \citet{mulleretal2018aa617_L2M}; $^{22}$ \citet{keppleretal2018aa617_A44}; $^{23}$ \citet{haffertetal2019natas3_749}; $^{24}$ \citet{currieetal2023sci380_198}; $^{25}$ \citet{mesaetal2023aa672_a93}; $^{26}$  \citet{derosaetal2023aa672_A94}; \mbox{$^{27}$ \citet{fransonetal2023apj950_L19}; } $^{28}$ \citet{Matthewsetal2024naturepreview}.}} 
\label{tab:remarkableB}
\end{table}%



\subsection{GQ Lup~b}
\label{sec:gqlup}
\begin{figure}[H]
\includegraphics[width=11.5 cm]{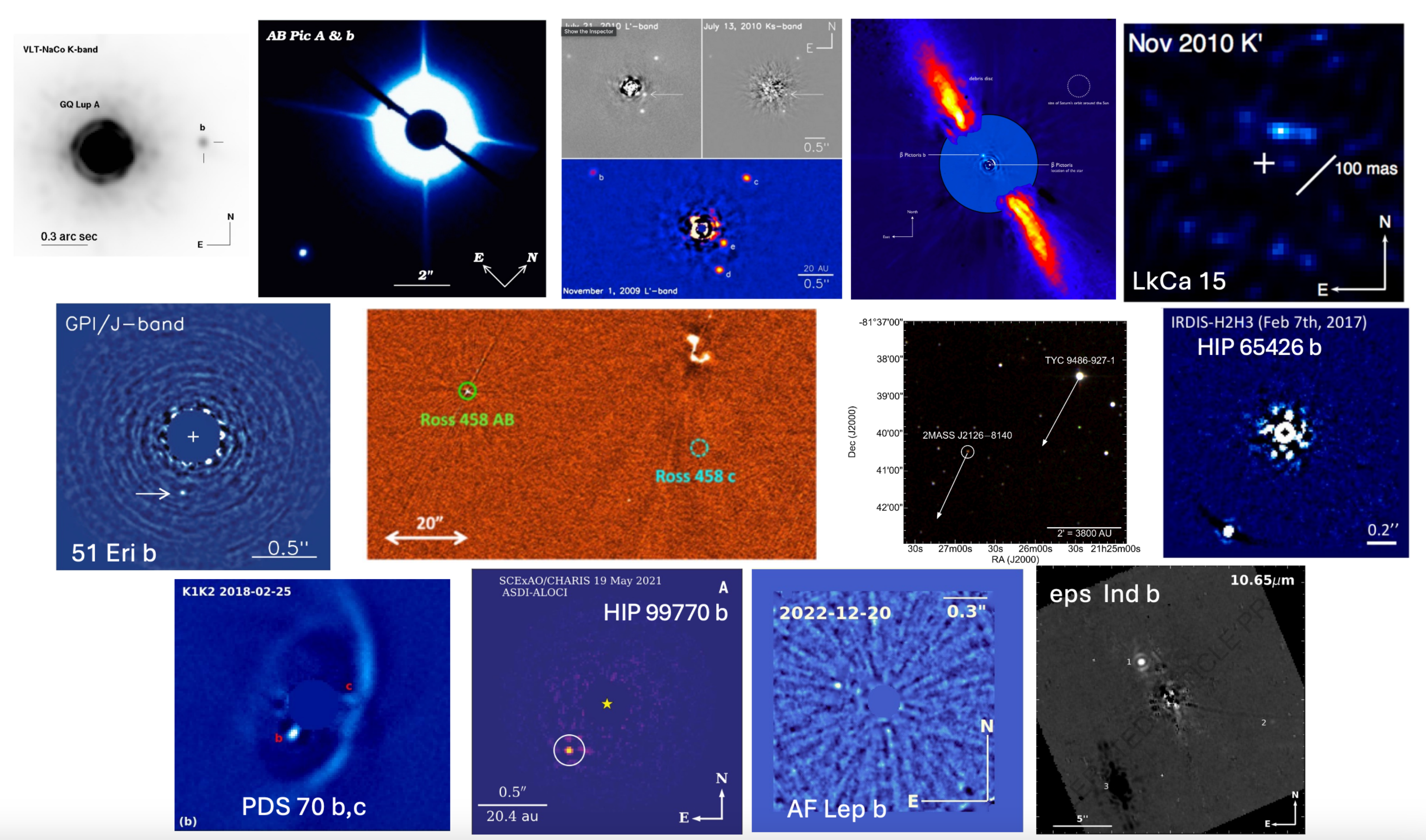}
\caption{
Discovering images of all the planets described in the Section\ \ref{sec:res} and the reference of each discovery paper.
Starting from left top to bottom right, there is the following: GQ Lup \citep{neuhauseretal2005aa435_L13}; \mbox{AB Pic A b \citep{chauvinetal2005aa438_L29};} \mbox{HR 8799 b,c,d,e \citep{maroisetal2010Natur468_1080};} $\beta$ Pic b \citep{lagrangeetal2009aa493_L21}; Ross 458 (AB) b (VLA-C band images \citep{cendesetal2022aj163_15}); LkCa 15 b \citep{krausandireland2012apj745_5}; \mbox{51 Eri b \citep{macintoshetal2015sci350_64};}  2MJ2126 \citep{deaconetal2016mnras457_3191}; HIP 66426 \citep{chauvinetal2017aa605_9}; PDS 70 \citep{2019A&A...632A..25M}; HIP 99770 b \citep{currieetal2023sci380_198}; AF Lep b \citep{mesaetal2023aa672_a93, derosaetal2023aa672_A94, fransonetal2023apj950_L19}; $\epsilon$ Ind b \citep{Matthewsetal2024naturepreview}.}
\label{fig:familyportrait}
\end{figure}

\textls[-20]{The classical T Tauri star GQ Lup (K7eV) in the Lupus star-forming region (\mbox{D = $140 \pm 50$ pc,} age $\leq 2$ Myr) is the primary of a wide binary system (a = 2400 au, \citep{alcalaetal2020aa635_L1}) with the secondary, GQ Lup B with M$\sim 0.15$\ M$_\odot$. In~2005 around the GQ Lup A, \citet{neuhauseretal2005aa435_L13} discovered a low mass companion using VLT/NACO (see Figure\ \ref{fig:familyportrait}). The~companion, 6 mag fainter than GQ Lup A, has a semi-major axis of $32.0 \pm 2.25$ au (\mbox{angular separation = 0.7 arcsec}).
The mass of GQ Lup b has been debated ever since its discovery, with~inferred masses of $\sim$10--40 M$_\text{J}$ (e.g., \citep{maroisetal2007apj654_l151, mcelwainetal2007apj656_505, seifhartetal2007aa463_309}). GQ Lup A has a circumstellar disk with a gap detected at $\sim$10 au, which could be the evidence of a hidden planet on a solar-system scale  \citep{longetal2020apj895_L46}.}

\subsection{AB Pic~b}
\label{sec:abpic}
In 2005, \citet{chauvinetal2005aa438_L29} using both the Adonis/SHARPII@ESO3.6 telescope and VLT/NACO found a companion orbiting the young K2V star ($\sim$30 Myr) member of the large Tucane-Horologium association (see Figure\ \ref{fig:familyportrait}). The~companion has been detected in the J, H, and~K bands at an angular separation of 5.5 arcsec. Comparing the photometry of AB Pic b with theoretical models \citep{burrowsetal1997apj491_856, chabrieretal2000apj542_464, baraffeetal2002aa382_563B} the authors stated a mass of about 13--14 M$_J$ and a value for the temperature ranging between 1513 and 1856 K allowing the authors to derive a spectral type L$1^{+2}_{-1}$. This object was the first object observed with the HCI technique with the mass at the deuterium burning limit. Is it a planet or a BD? The object became part of a debate relating to the two criteria of the definition of a planet (the mass limit of \mbox{13 M$_J$} or the formation scenario). Following the formation scenario criterion,  the~ apparent absence of companions of low masses in the results of the major surveys suggests that wide companions like AB Pic b ($\sim$260 au) have formed by gravitational instability \citep{chauvinetal2005aa438_L29}. In~a further analysis of AB Pic b observations with SPHERE, \citet{palmabifanietal2023aa670_a90} confirmed the AB Pic b parameters and found an obliquity of the planet between 45 and 135 deg, which is evidence of a misalignment between the stellar spin and the planetary orbit. Moreover, by mixing the HARPS-N radial velocity measurements by \citet{grandjeanetal2020aa633_A44} with the proper motion anomaly, these authors suggested the presence of a second inner planet (AB Pic c) with a mass of $\sim$6 $\text{M}_J$ orbiting between 2.5 and 10~au.

\subsection{HR 8799 b,c,d,e}
\label{sec:hr8799}

The huge forward leap in HCI was the discovery of the first and for many years the only multi planetary system using direct imaging techniques with Keck and Gemini telescopes around the star HR\,8799 \citep{maroisetal2008sci322_1348, maroisetal2010Natur468_1080} (see Figure\ \ref{fig:familyportrait}).
The host star is an A5 spectral type with a mass of around 1.5~$M_{\odot}$ and an estimated age of around 30~Myr. The~star is at a distance of 39.4~pc from the Sun, and it is proposed to be part of the Columba moving group \citep{2008hsf2.book..757T}. The~three planets {\it b}, {\it c}, and {\it d} were discovered in 2008 at a separation of 1.70$"$, 0.94$"$, and 0.64$"$, which correspond to projected separations of $\sim$68~au, $\sim$38~au, and  $\sim$24~au, respectively. Later, in~2010, a~fourth inner planet, HR\,8799\,e, was discovered at a separation of 0.37$"$, which corresponds to around 14~au. The~masses of these planets had a lot of different estimates also according to the very uncertain age of the system. Anyway, at~the moment, the most accepted values are in the range of 5--7~$M_{Jup}$. 
A wealth of different studies have been carried out in the following years both on the characteristics of every single planet and on the architecture
and the stability of the system (e.g., \citep{2010ApJ...710L..35J,2010ApJ...710.1408F,2011A&A...528A.134B, skemeretal2012apj753_14,marleyetal2012apj754_135,2013A&A...549A..52E,Ingrahametal2014apj794_L15, konopackyetal2016aj152_28, 2016A&A...587A..57Z,2016A&A...587A..58B,
greenbaumetal2018aj155_226,2018ApJS..238....6G,2020MNRAS.491.1795P,2020AJ....160..150W,2022A&A...666A.133Z}). \par
In particular, \citet{konopackyetal2016aj152_28}, after a long astrometric monitoring at Keck observatory, showed that the whole system is close to a 1:2:4:8 resonance and the b,c,d planets lay on a quite coplanar ($\sim$27$^\circ$) orbits with the disk. Instead, planet e appears not to be in a coplanar orbit with the b,c, and~d planets \citep{gravitycollaborationetal2019aa623_L11}. The~planets have wide separation from the host star, allowing extensive and wide wavelength range spectroscopy observation of the planets without contamination from the host star. \citet{skemeretal2012apj753_14} shown that all planets are  bright at 3.3 $\upmu$m compared with equilibrium chemistry model used for field BDs. \citet{oppenheimeretal2013apj768_24} were able to tentatively identify in all  the planets of the system CH$_4$, NH$_3$, C$_2$H$_2$ and possibly CO$_2$ or HCN, using Project1640 at Palomar Hale  5 m Telescope. The~interested reader could find other spectroscopical results in the following references: \citet{marleyetal2012apj754_135,Ingrahametal2014apj794_L15,greenbaumetal2018aj155_226,billeretal2021mnras503_743,wangetal2022aj164_143}.

 \subsection{$\beta$ Pic b,c}
 \label{sec:betapic}

Another important system is that of $\beta$\,Pic an A6 V star at a distance of 19.3~pc from the Sun and with an estimated age of $\sim$24~Myr. It hosts a well-studied edge-on dust disk that since its discovery \citep{1984Sci...226.1421S} was thought to be an ideal environment to host a planetary system (see Figure\ \ref{fig:familyportrait}). The~planetary mass companion was finally discovered at a separation of around 10~au and with an estimated mass just above 10~$M_{Jup}$ \citep{lagrangeetal2009aa493_L21}. Its relatively low orbital period of just above 20~years allowed for a good characterization of its orbit parameters (e.g., \citep{2012A&A...542A..41C}), and it also made it possible to evaluate its interaction with the debris \mbox{disk \citep{2012A&A...542A..40L}.} Moreover, coupling the astrometric data from Hipparcos and Gaia with those from \mbox{HCI \citep{2018NatAs...2..883S}} were able for the first time to obtain a model-independent estimate of the mass of $\beta$\,Pic\,b obtaining a value of 11 $\pm$ 2~$M_{Jup}$. The~orbit of the planet was further constrained thanks to the possibility of observing it before and after its conjunction \citep{2019A&A...621L...8L}. Finally, evidence for a second inner planet in the system ($\beta$ Pic c) was found through radial velocity data \citep{lagrangeetal2019NatAs3_1135}. From~astrometric  and RV data, the~two planet masses have been estimated as $9.3^{+2.5}_{-2.6}$\ M$_J$ for $\beta$ Pic b and $8.3 \pm 1$\ M$_J$  for $\beta$ Pic c . The latter was directly imaged by \citet{nowaketal2020aa642_L2} with high-contrast interferometric imaging by VLT/GRAVITY \citep{gravitycollaboration2020aa633_A110}. $\beta$ Pic c has been the first planet found with the RV technique and confirmed with~DI.

\subsection{Ross 458 (AB) b}
\label{sec:ross458}
The binary system Ross 458 (AB) consits of a M0.5 primary and a M7 secondary orbiting at 5.4 au (in February 2000, \citep{beuzitetal2004aa425_997}) with a period of 14.5 yr (see Figure\ \ref{fig:familyportrait}). In~2010, \citet{goldmanetal2010mnras405_1140} identified a wide orbit companion (1168.0 au) of Ross 458 (AB) with GROND (Gamma-ray Burst Optical/Near-infrared Detector) at the 2.2 m ESO/MPG telescope and Omega 2000 NIR wide-field imager on the Calar Alto 3.5-m telescope.  Ross 458 (AB) b shares its proper motion with the binary, forming, in~this way, a~hierarchical low-mass star plus PMC system. The~authors estimated an age $\leq 1$\ Gyr and M$\sim$14 M$_J$ identifying Ross 458 (AB) b as the first circumbinary companion with a mass close to the deuterium burning limit ever discovered. Further observations of Ross 458 (AB) b show the presence of clouds (e.g., \citep{morleyetal2012apj756_172} and references therein) that are opacity sources expected from the condensation of species at the temperature of these sub-stellar objects (\mbox{$650 \pm 25$\ K \citep{burgasseretal2010apj725_1405}}). It is worth noting that Ross (AB) b is another object under discussion for its belonging, or~not, to~the realm of planets. Its mass is just close to the deuterium burning limit, its surface temperature is lower that that of well know planets (e.g., AB Pic b, HR 8799 b). However, its wide separation raises doubts about its planetary character \citep{burgasseretal2010apj725_1405}.

\subsection{Lk\,Ca\,15 b,c}
\label{sec:lkca}

The central star of Lk\,Ca\,15 is a K5 spectral type T-Tauri star with an age of $\sim$2~Myr and at a distance of around 157\ pc from the Sun. The~star is known to host a transitional proto-planetary disk \citep{2006A&A...460L..43P,2008ApJ...682L.125E}, and it is thought to be an ideal place for the search for planetary-mass companions caught during their formation. Different searches for these companions were performed since the discovery of the disk (e.g., \citep{2010A&A...522A...2B}), and the first detection of a possible companion was for the first time proposed \mbox{by \citet{krausandireland2012apj745_5}} at a separation of $71.9 \pm 1.6$\ mas, corresponding to a projected separation of $\sim$20\ au from the host star well into the gap of the disk (see Figure\ \ref{fig:familyportrait}). Later, the~presence of three accreting companions embedded in the disk was also proposed  by \citet{sallumetal2015Natur527_342}. The~nature of such objects was, however, widely disputed, with other authors proposing that such structures could have been explained with scattered light by extended structures of the disk \citep{thalmannetal2016ApJ828_L17, mendigutiaetal2018aa618_L9, blakelyetal2022apj931_3, sallumetal2023apj953_55}.
The presence of possible companions embedded in the disk of such a star is still strongly debated, and its case could be taken as an example of the extreme difficulty in directly imaging young giant planets embedded in proto-planetary disks. 

\subsection{51 Eri~b}
\label{sec:51eri}

A further step forward was obtained with the beginning of the operativity of the new generation of high-contrast imagers like GPI and SPHERE in 2014 and 2015
that allowed both the detection of new sub-stellar companions and a better characterization of those that were previously known. The~first companion
detected using one of these new instruments with extreme adaptive optics modules was 51\,Eri\,b (GPI, \citep{macintoshetal2015sci350_64}). In~this case the host star is a F0 star with a mass of 1.75~$M_{\odot}$ and at a distance of 29.4~pc from the Sun. The~companion was detected at a separation of $\sim$0.450$"$, which corresponded to a projected physical separation of 
around 13~au (see Figure\ \ref{fig:familyportrait}). Its mass was estimated to be $\sim$2~$M_{Jup}$, assuming for the system an age of 20~Myr. \citet{macintoshetal2015sci350_64} also reported spectral observation made with NIRC2 at Keck telescope, with the spectra recalling those of a field brown dwarf of spectral type T4.5 to T6, with~the presence of strong methane absorption, similar to that of~Jupiter.

\subsection{2MJ2126}
\label{sec:2mj2126}
\textls[-15]{ Short for 2MASS J21265040?8140293, 2MJ2126, formerly identified as standalone \mbox{object \citep{fahertyetal2013aj145_2},}  is the fainter (J = 15.5 mag) component of a very wide separation (\mbox{$\sim$217 arcsec}) common proper motion system \citep{deaconetal2016mnras457_3191} (see Figure\ \ref{fig:familyportrait}). The~brighter companion (\mbox{J = 8.2 mag}) is TYC 9486-927-1, an~active, rapidly rotating early (1--45 Myr) M1 dwarf \mbox{\citep{torresetal2006aa460_695}. \citet{deaconetal2016mnras457_3191}} evaluated the 2MJ2126's mass ranging from 11.6 to 15.0 M$_J$, using theoretical models by \citet{baraffeetal2003aa402_701}, placing this object on the deuterium burning threshold. Actually, 2MJ2126  is not one of the results obtained with the HCI, but~it is likely the widest orbit planetary-mass object known (>4500 au), and its estimated mass, age, spectral type, and~$\teff$ are similar to the well-studied planet $\beta$ Pictoris b \citep{deaconetal2016mnras457_3191}.}

\subsection{HIP 65426~b}
\label{sec:hip65426}
\citet{chauvinetal2017aa605_9} discovered the first SPHERE sub-stellar companion around the 17~Myr old, A2 spectral type star HIP\,65426 at a distance of 111.4~pc from the Sun. The~companion, HIP\,65426\,b, was detected at a separation of 0.83$"$, which corresponded to a projected separation of 92~au, and~such a large separation, together with the unusually large rotation rate of the central star, suggested a planet-planet scattering mechanism with the engulfment of one of the two planets (see Figure\ \ref{fig:familyportrait}). The~mass of the companion was estimated to be of the order of 6--12~$M_{Jup}$. The~analysis of the IFS spectra confirms a low-surface-gravity atmosphere of spectral type L6 $\pm$ 1 for HIP 65426 b consistent with a young massive planet at the age of the Lower Centaurus--Crux association.
Recently, the~ERS program high-contrast Imaging of Exoplanets and Exoplanetary Systems with JWST (ERS-01386; \citep{hinkleyetal2022pasp134_5003}) tested the high-contrast exoplanet imaging of both NIRCam and MIRI and~coronagraphic imaging ($2\leq \lambda \leq$ 16 $\upmu$m) modes of JWST, observing exoplanet HIP 65426b \citep{carteretal2023ApJ951_L20}. HIP 65426b has been clearly detected in all seven observational filters (F250M, F300M, F356M, F356W, F410M, F444W, F1140C, F1550C), and~it is the first-ever exoplanet detected beyond 5 $\upmu$m.

\subsection{PDS 70 b,c}
\label{sec:pds70}

PDS\,70 is a K7 star with a mass 0.76~$M_{\odot}$ and at a distance of 112~pc from the Sun. The~estimated age of the system is of the order of 5~Myr. This star is known to host a proto-planetary disk with a gap with a radius of the order of 70~au \citep{2012ApJ...758L..19H}. The~first companion was detected at a separation of 0.195" corresponding to a projected separation of the order of 22~au well into the gap in the disk (see Figure\ \ref{fig:familyportrait}). The~mass of the companion was estimated in several works ranging from  <1 $M_{Jup}$ to more than 10~$M_{Jup}$ (e.g., \citep{mulleretal2018aa617_L2M,2019A&A...632A..25M,2020AJ....159..263W,2020A&A...644A..13S}).  PDS 70 b is on eccentric orbit (e $\sim 0.17 \pm 0.06$).
Follow-up observations in $H\alpha$ narrow band using MUSE allowed the detection of a second accreting planet (PDS\,70\,c; \citep{haffertetal2019natas3_749}) at a separation of $\sim$0.250$"$, which corresponds to around 34~au from the host star. This second companion was not detected at longer wavelengths because it was partially hidden by the disk, but once detected, it was possible to recover it also in NIR with SPHERE \citep{2019A&A...632A..25M}, allowing its characterization with an estimated mass of the order of 4.5~$M_{Jup}$. In contrast to PDS 70 b, PDS 70 c shows a circular orbit.
Finally, another important result was the detection for the first time of a circum-planetary disk around PDS\,70\,c using ALMA at sub-millimeter wavelengths with an extension of around 1.2~au \citep{2021ApJ...916L...2B} and an estimated mass of 0.007--0.03 $M_\oplus$. While, as~we have seen above, other systems have been proposed. This is, up to now, the only system for which the presence of directly imaged companions has been proved in the cavity of a proto-planetary disk. This is, of course, of paramount importance as it allows us to validate models for which the presence of structures as gaps and spirals are linked to the presence of forming planets  embedded in the disks~themselves. \par


\subsection{HIP 99770~b}
\label{sec:hip99770}
HIP 99770b could be considered the first planet found with the PMa technique. The~discovery of this object has been made with the coupling of SCExAO/CHARIS and Keck/NIRC2 data with Gaia and Hypparcos astrometry \citep{currieetal2023sci380_198}. The~host star is an A spectral type with an age ranging between 40 Myr and 414 Myr as determined by Tess Asteroseismology. HIP 99770 b, orbiting 17 au from its star, has a dynamical mass of 13.9--16.9  M$_\text{J}$ and a planet-to-star mass ratio $\sim 8 \times 10^{-3}$ similar to other planets discovered with DI \citep{currieetal2023sci380_198}.

\subsection{AF Lep~b}
\label{sec:aflepb}
AF Lep is an F8 star with a mass of 1.2~$M_{\odot}$ and at a distance of 26.8~pc from the Sun. It is part of the $\beta$\,Pic moving group with an estimated age of 24 $\pm$3~Myr (\citep{mesaetal2023aa672_a93} and reference therein). AF Lep was classified as an SB2 star (spectroscopic binary) with a stellar companion at a separation of 0.021 au \citep{nordstrometal2004aa418_989, ekeretal2008mnras389_1722}; this classification is questioned by further analysis by \citet{mesaetal2023aa672_a93}. Furthermore, it is known to host a debris disk belt with a radius of \mbox{54 $\pm$ 6~au.} Using the PMa technique \citet{mesaetal2023aa672_a93, derosaetal2023aa672_A94, fransonetal2023apj950_L19} discovered the presence of the PMC AF\,Lep\,b (see Figure\ \ref{fig:familyportrait}). The~2--5~$M_{Jup}$ planet is orbiting its host star at a separation of 8--9~au. The~analysis of the IFS spectra shows that AF Lep b is a late-L spectral type object with a $\teff$ ranging between 1000 and 1700 K. This is the first companion with a mass below the deuterium burning limit discovered by coupling direct imaging with PMa measurements. Moreover, it could be considered the first directly detected Jupiter analog orbiting an accelerating star \citep{mesaetal2023aa672_a93}.

\subsection{$\epsilon$ Ind~b}
\label{sec:epsind}
$\epsilon$ Ind  is a triple system composed of a K2V (M = 0.76 M$_\odot$) star ($\epsilon$ Ind A) with an age of $\sim 3.5$\ Gyr and~a binary system with two BDs: one ($\epsilon$ Ind B) is a T1.5 BD of M = 75.0 M$_J$ and the companion ($\epsilon$ Ind C) is a T6 BD with a mass of M = 70 M$_J$. The~two BDs are separated by a semimajor axis of 2.6 au while they are separated by $\epsilon$ Ind A by a projected distance of 1460 au \citep{philipotetal2023aa670_A65}. $\epsilon$ Ind A b was detected by \citet{fengetal2019mnras490_5002} combining RV data and Gaia astrometric data. Further, \citet{philipotetal2023aa670_A65} revised the orbital parameters of $\epsilon$ Ind A b evaluated by \citet{fengetal2019mnras490_5002}, finding very different values, with~M = 3.0 M$_J$, a~semimajor axis of 8.8 au and an eccentricity of 0.48. Observing $\epsilon$ Ind A with MIRI and a coronagraphy mode of JWST, \citet{Matthewsetal2024naturepreview} revealed a giant planet  (see Figure\ \ref{fig:familyportrait}) which is consistent with the astrometric and RV measurement reported previously. Using both the RV measurement of the object, the~astrometric measurement, and the images obtained by JWST, \citet{Matthewsetal2024naturepreview} constrained the orbital parameter with a M = $6.31^{+0.60}_{-0.56}$\ M$_J$, a~semi-major axis of $28.4^{+10.0}_{-7.2}$\ au. The~orbit has 
an eccentricity of $0.40^{+0.15}_{-0.18}$. Also, if the latter orbital parameter is in disagreement with the findings in \citet{philipotetal2023aa670_A65}, \citet{Matthewsetal2024naturepreview} concluded that the data indicate that it is the only giant planet in the system and that it could be referred as b. The~atmospheric modeling of the mid NIR photometry is consistent with a planet of T$\sim 300$\ K, which makes $\epsilon$ Ind A b not only the first planet discovered by JWST but~also the coldest planet directly~imaged.

\startlandscape
\begin{table}[H]
\caption{Deep imaging Surveys dedicated to the search for planets around young and intermediate-age stars. Table adapted from \citet{chauvinetal2015aa573_A127, chauvin2018SPIE10703E05} and updated. We have indicated the telescope, the~instrument, the~imaging mode (Cor-I: coronagraphic imaging; Sat-I; saturated imaging; I: imaging; SDI: simultaneous differential imaging; ADI: angular differential imaging; ASDI: angular and spectral differential imaging; APP-ADI: Apodizing mask angular differential imaging), the~filters, the~field of view (FoV), the~number of stars observed (\#), their spectral types (SpT), and~ages (Age). }\label{tab:revhcisurvey2}
\newcolumntype{C}{>{\centering\arraybackslash}X}
\newcolumntype{L}{>{\raggedright\arraybackslash}X} 
\begin{tabularx}{\textwidth}{LLCCCCCCCC}
\toprule
\textbf{Survey}   & \textbf{Tel.}   & \textbf{Instr.}    & \textbf{Mode}  & \textbf{Filter}    & \textbf{FoV}                &   \textbf{\# }   &  \textbf{SpT}   &\textbf{Age }                  &  \textbf{Ref. }\\
              &          &             &            &             &\textbf{(arcsec\boldmath{$^2$})}   &          &            &\textbf{ (Myr)}               &          \\
\midrule

NA94                                & PALOMAR       & AOC       & Cor-I  & I          & $60 \times 60$ & 24    & G-M   & FIELD &  \cite{nakajimaetal1994ApJ428_797}\\
... \\
CH03                                 & ESO3.6m         & ADONIS & Cor-I  & H,K
      &$13 \times 13$ & 29    & GKM  &$\lesssim$50   &  \cite{Chauvinetal2003aa404_157}      \\
NE03                                 & NTT                  & Sharp     & Sat-I   & K         & $11 \times 11$ & 23    & AFGKM & $\lesssim$50   &  \cite{neuhauseretal2003an324_535}      \\
                                          & NTT                  & Sofi        & Sat-I   & H         & $13 \times 13$ & 10    & AFGKM & $\lesssim$50   &  $-$ \\
LO05                                 & HST                  & NICMOS &Cor-I   & H         &$19 \times 19$ & 145   & AFGKM & 10--600   &  \cite{lowranceetal2005aj130_1845}      \\
HMPES $^{a}$  &  VLT                  & NaCo    & Sat-I   & H,K     &$14 \times 14$ &  28    &  KM    &$\lesssim$200 &  \cite{masciadrietal2005apj625_1004}      \\
BI07                                  &  VLT                   & NaCo    & SDI    & H         &$5 \times 5$ &  45    &  GKM    &$\lesssim$300 &  \cite{billeretal2007apjs173_143}      \\
                                         &  MMT                   &             & SDI    & H         &$5 \times 5$ &  $-$    &  $-$    &$-$ &  $-$      \\
KA07                                &  VLT                    & NaCo     &Sat-I   & L$'$        & $28 \times 28$ &  22    &  GKM    &$\lesssim$50 &   \cite{kasperetal2007aa472_321}      \\
GDPS $^{b}$   &  Gemini - N          & NIRI       & ADI      &H      &$22 \times 22$ &  85    &                & 10--5000 &  \cite{lafreniereetal2007apj_670_1367}\\ 
AP08                                & VLT                     & NaCo      &SDI     & H      & $3 \times 3$    &  8       &  FG        &12--500    & \cite{Apaietal2008apj672_1196}  \\
CH10                                & VLT                     & NaCo      & Cor-I  & H,K   & $28 \times 28$ &  88    &  BAFGKM    &$\lesssim$100 &  \cite{chauvinetal2010aa_509_A52}      \\
HE10                                & MMT                   & Clio         & ADI     &L$'$, M  & $15.5 \times 12.4$ &  54    &  FGK    &100--5000 & \cite{heinzeetal2010apj714_1551, heinzeetal2010apj714_1570}      \\
JA11                                 & Gemini - N          & NIRI        & ADI     & H,K   & $22 \times 22$      &  15     &   BA      & 20--700    & \cite{jansonetal2011apj736_89}   \\
IDPS $^{c}$      &  GEMINI-N        & NIRI      & ADI      &J,H,K    &$22 \times 22$ &  292    &  BAFGKM & <1000 &  \cite{viganetal2012aa544_A9, Galicheretal2016aa594_A63}\\  
                                          & KECK II             & NIRC2  & ADI     & J,H,K,L  &$10 \times 10$ &    $-$        & $-$   & $-$   & $-$   \\
                                          & GEMINI-S         & NICI      & ADI /SDI    & H,K  &$18\times 18$ & $-$   & $-$   & $-$   & $-$   \\ 
                                          & VLT                   & NaCo     & ADI     & H,K        &$14 \times 14$ & $-$   & $-$   & $-$   & $-$   \\
DE12                                 & VLT                   & NaCo      &ADI     & L$'$           & $28 \times 28$ & 16     &  M    & $\lesssim$200 &  \cite{delormeetal2012aa539_A72}      \\
PALMS $^{d}$  & KECK II             & NIRC2   & ADI      & H,K       &  $10 \times 10$ & 72  &  M                      & $\lesssim$300 & \cite{bowleretal2012apj753_142, bowleretal2012apj756_69, bowleretal2015apjs216_7}  \\
                                         & SUBARU          & HiCIAO  &ADI      & H,K       & $20 \times 20$ & $-$   & $-$   & $-$ &  $-$  \\
RA13                                 & VLT                   & NaCo      & ADI    & L$'$           & $28 \times 28$ & 59     &  AF    & $\lesssim$200 & \cite{rameauetal2013aa553_A60}     \\
SEEDS $^{e}$    & SUBARU           & HiCIAO  &ADI      & H,K       & $20 \times 20$ & 20    & AFGKM            & $\lesssim$1000 & \cite{tamura2009AIPC1158_11, yamamotoetal2013PASJ65_90, brandtetal2014apj794_159, jansonetal2013apj773_73, tamura2016PJAB92_45}\\
BI13                                  & Gemini-S           & NICI      & Cor-ASDI& H       & $18 \times 18$ & 80    & BAFGKM         & $\lesssim$200 & \cite{billeretal2013apj777_160}\\
NI13                                  & Gemini-S           & NICI      & Cor-ASDI& H       & $18 \times 18$ & 70    & BA                   & 50--500 & \cite{nielsenetal2013apj776_4}\\
WA13                                & Gemini-S           & NICI      & Cor-ASDI& H       & $18 \times 18$ & 57    & AFGKM           & $\lesssim$1000 & \cite{Wahhajetal2013apj773_179}\\
LEECH $^{f}$   & LBT                    & LMIRCam& ADI      & H,L$'$    & $11 \times 11$ & 136   & AFGKM           & $\lesssim$1000 & \cite{skemeretal2014SPIE9148E_0L, maireetal2015AA579_C2, stoneetal2018aj156_286} \\
SPOTS $^{g} $ &  VLT                   &        NaCo & ADI     &   H      &$14 \times 14$ &26     &AFGKM             &$\lesssim$200       &\cite{thalmannetal2014aa572_A91, bonavitaetal2016aa593_A38, asenioTorresetal2018aa619_A43}       \\
                                          & $-$                     &SPH/IRDIFS (IFS)       & ADI        & Y,J        &   $1.77 \times 1.77$& 34 & $-$  &$-$  & $-$  \\
                                          & $-$                     & SPH/IRDIFS (IRDIS)   & ADI        & H          &   $11 \times 11$      & $-$ & $-$ &  $-$  &  $-$\\        
ME15                                 &VLT                     & NaCo       & APP-ADI& L$'$     &$28 \times 28$  & 20     &AF                   & $\lesssim$200  &\cite{meshkatetal2015apj800_5, meshkatetal2015MNRAS453_2533}\\ 
DU16                                 & Spitzer               & IRAC       &    I            &4.5 $\upmu$m   & $312 \times 312$& 73  &AFGKM &   $\lesssim$200  & \cite{Durkanetal2016apj824_58}\\
NACO LP $^{h}$ & VLT                   & NaCo    & ADI     & H       &$14 \times 14$ &  86    &  FGK & $\lesssim$200 &  \cite{desideraetal2015aa573_A126, chauvinetal2015aa573_A127, viganetal2017aa603_A3}\\  
MASSIVE $^{i} $& VLT                   & NaCo    & ADI     & L$'$       &$28 \times 28$ &  58    &  M & $\lesssim$100 &  \cite{lannieretal2016aa596_A83}\\  
PSYM-WIDE $^{j}$ & Gemini-S     & GMOS-S  & CI   & i$'$,z$'$     & $300 \times 300$&95    & KML & $\lesssim$200  & \cite{naudetal2017aj154_129} \\
\bottomrule
\end{tabularx}
\end{table}

\begin{table}[H]\ContinuedFloat
\caption{{\em Cont.}}
\newcolumntype{C}{>{\centering\arraybackslash}X}
\newcolumntype{L}{>{\raggedright\arraybackslash}X} 
\begin{tabularx}{\textwidth}{LLCCCCCCCC}
\toprule
\textbf{Survey}   & \textbf{Tel.}   & \textbf{Instr.}    & \textbf{Mode}  & \textbf{Filter}    & \textbf{FoV}                &   \textbf{\# }   &  \textbf{SpT}   &\textbf{Age }                  &  \textbf{Ref. }\\
              &          &             &            &             &\textbf{(arcsec\boldmath{$^2$})}   &          &            &\textbf{ (Myr)}               &          \\
\midrule

WEIRD $^{k}$                                &Spitzer               & IRAC       &    I            &3.6\ \& 4.5 $\upmu$m   & $312 \times 312$&344 &BAFGKM & 10--150 &\cite{baronetal2018aj156_137}\\
                                            &CFHT                  &WIRCam & CI            & J         & $1200 \times 1200$ & $-$        & $-$        & $-$\\
                                            &CFHT                  &MEGACam & CI            & z        & $3600 \times 3600$ & $-$        & $-$        & $-$\\
                                            &Gemini-S           &GMOS-S & CI            & z         & $300 \times 300$ & $-$        & $-$        & $-$\\
                                            &Gemini-S           &Flamingos-2 & CI            & J         & $360 \times 360$ & $-$        & $-$        & $-$\\
BEAST $^{l}$   & VLT                  &SPH/IRDIFS (IFS)       & ADI        & Y,J,H &   $1.77 \times 1.77$& 85 & B  &10--20  & \cite{jansonetal2019aa626_A99, Jansonetal2021aa646_A164} \\
                                           & $-$                   & SPH/IRDIFS (IRDIS)   & ADI        & H,K          &   $11 \times 11$      & $-$ & $-$ &  $-$  &  $-$\\  
YSES $^{m}$      & VLT                   & SPH/IRDIS   & CI/DBI        & Y,J,H,K          &   $11 \times 11$      & 70 &  K &  $\sim$ 15  &  \cite{bohnetal2020mnras492_431, bohnetal2020apjl, bohnetal2021aa648_A73}\\ 
                                          & $-$                    &NaCo                    &   CI             & L$'$,M$'$              &   $28 \times 28$      & $-$  &$-$   & $-$ & $-$ \\   
GPIES $^{n}$     &Gemini-s           & GPI       & ADI/SDI &Y,J,H,K  & $2.78 \times 2.78$ & 600& BAFGKM& $\lesssim$1000  & \cite{nielsenetal2019aj158_13}\\
SHINE $^{o}$    & VLT                  &  SPH/IRDIFS (IFS)       & ADI        & Y,J        &   $1.77 \times 1.77$& 500& AFGKM&  $\lesssim$800   & \cite{desideraetal2021aa651_a70, viganetal2021aa651_a72, langloisetal2021aa651_a71} \\
                                           &          $-$          & SPH/IRDIFS (IRDIS)   & ADI        & H          &   $11 \times 11$      & $-$ & $-$ &  $-$  &  $-$\\
ISPY $^{p}$      &VLT                     & NaCo       & ADI& L$'$     &$28 \times 28$  & $\sim 200$     &BAFGKM  & <10 &\cite{launhardtetal2020aa635_A162}\\
 \\ 
\midrule
\end{tabularx}

\begin{adjustwidth}{+\extralength}{0cm}
 \noindent{\footnotesize{Notes: 
 $^\text{a}$ Hot Massive Extrasolar Planets Search; $^\text{b}$ Gemini Deep Planet Survey;  $^\text{c}$ International Deep Planet Survey;  $^\text{d}$ Planets Around Low-Mass Stars; $^\text{e}$ Strategic Exploration of Exoplanets and disks;  $^\text{f}$ LBTI Exozodi Exoplanet Common Hunt ;  $^\text{g}$ Search for Planets Orbiting Two Stars; $^\text{h}$ NaCo Long Program 184.C-0157;  $^\text{i}$ M-dwArf Statistical Survey for direct Imaging of massiVe Exoplanets; $^\text{j}$ Planet Search around Young-associations; $^\text{k}$ Wide-orbit Exoplanet search with Infrared Direct imaging; $^\text{l}$ B-star Exoplanets Abundance Study; $^\text{m}$ Young Suns Exoplanets Survey;  $^\text{n}$ Gemini Planets Imager Exoplanet Survey;  $^\text{o}$ SpHere INfrared survey for Exoplanets; $^\text{p}$ Imaging Survey for Planets around Young Stars.}}
\end{adjustwidth}
\end{table}%
\finishlandscape

\section{Occurrence Rate of Low Mass Companions on Wide~orbits}
\label{sec:statistics}
One of the main goals of large direct-imaging surveys (see Table\ \ref{tab:revhcisurvey2}) is the statistical definition of the frequency of exoplanets as a function of their separation from the host star. In~the following we describe some of the main surveys and their results.
The~first example is, perhaps, the survey performed in the L$'$ spectral band using NACO of 22 targets in the Tucana and $\beta$-Pictoris moving groups (KA07, \cite{kasperetal2007aa472_321}) that, despite the lack of novel detection, costrained the distribution of Jupiter-like companions with M > 3 $\text{M}_\text{J}$ at separation larger than 30 au with a lower frequency (<2\%) than that stated by the HMPES (Hot Massive Extrasolar Planets Search) results. In~the following years, new and larger surveys were conducted to allow for the refinement of these results.
\citet{nielsenandclose2010apj717_878} exploited a survey on 118 young FGKM stars observed both with NACO in H and K spectral bands and the NIRI camera at the Gemini telescope (HMPES \& BI07) and using different planetary models to infer the frequency of low mass companions. They were able to set an upper limit of $\sim$20\% for the presence of planetary mass companions ($\geq$4 $\text{M}_\text{J}$) around these stars and to put a model-depending limit for the separation beyond which the presence of a planet was not possible. \citet{heinzeetal2010apj714_1570} (HE10) prioritized the nearness of the sample selecting 54 F, G and K stars at a median distance of 11.2 pc from the Sun. Independently by distribution models, they found with a confidence of 90\% that less than 50\% of FGK stars closer than 11.2 pc have a planetary companion with $\text{M}> 5\text{M}_{J}$ orbiting between 30 and 94 au. \citet{jansonetal2011apj736_89} (JA11) observed 18 nearby  and very massive stars with spectral type between B2 and A0 and~found that <30\% of these stars form and retain planets, BD, and~very low mass stars of $\text{M}<100\ \text{M}_\text{J}$ within 300 au, at~99\% confidence. Their conclusion took them to privilege the core accretion as the main planet formation mechanism around this type of star. 
The International Deep Planet Survey (IDPS, \cite{viganetal2012aa544_A9}) observed 54 young A and F spectral types stars at a small distance from the Sun using both NACO@VLT and NIRI@Gemini. They were able to define a frequency of 5.9--18.8\% for planets with mass between 3 and 14 M$_\text{J}$ at separations between 10 and 300 au. 
The Gemini-NICI planet-finding campaign (BI13, \cite{billeretal2013apj777_160}) was devoted to searching for low-mass companions around 80 members of the beta Pic, TW Hya, Tucana--Horologium, AB Dor, and Hercules--Lyra moving groups. They found a frequency lower than 18\% or 6\% according to the adopted model for planets in the mass range of 1--20 M$_\text{J}$ and separation of 10--150 au. The authors in \cite{rameauetal2013aa553_A60} (RA13) performed a survey of 59 A and F spectral type stars and found a frequency between 10.8\% and 24.8\% for planets with masses in between 1 and 13 M$_\text{J}$ and separation between 1 and 1000 au around these stars. 
A statistical analysis of different surveys (SEEDS, GDPS, and~BI13) was performed by \citet{brandtetal2014apj794_159}, and they found a low frequency of 1--3.1\% for companions with masses between 5 and 70 M$_\text{J}$ and for separations from the host star between 10 and 100 au. A~new \mbox{analysis \citep{Galicheretal2016aa594_A63}} of IDPS sample, considering a uniform distribution in planet masses and semi-major axis, demonstrated a frequency much lower than those found by previous survey with a frequency $\sim$1.05\%. If~considering a power law distribution, the frequency arrives at 2.3\% for planets with a mass between 0.5 and 14 M$_\text{J}$ and separations between 20 and 300 au. They also found no dependence for the frequency with the mass of the host star while this dependence is present for planets at smaller separation than that considered in their work.
A relatively less explored low mass stellar range was considered by \citet{lannieretal2016aa596_A83} (MASSIVE) that observed 54 M spectral type stars finding a frequency of 2.3\% for planetary mass companions at separations 8 and 400 au. They concluded that the low-mass stars host a different planetary mass population with respect to higher-mass host stars (A-F spectral types). 
The analysis of the results of the LEECH survey \citep{stoneetal2018aj156_286} considering 98 nearby stars with spectral types between B and M cannot exclude that planets with masses around 10 M$_\text{J}$ are quite common around FGK stars. 
An analysis of a much larger sample of confirmed members of nearby young moving groups was performed by  \mbox{\citet{baronetal2019aj158_187}} (WEIRD, and~PSYM-WIDE). Their sample was composed of 344 stars and considered companions with masses between 1 and 20 M$_\text{J}$ and separations between 5 and 5000 au, with a low frequency of  0.1\% in the case of the hot start model. Regarding the cold start model, they found, for the same semi-major axis and mass intervals, an~upper limit of 5.2\% at a 95\% confidence level. They also found that the occurrence of low-mass companions is negatively correlated with the orbit semi-major axis and positively correlated with the mass of the host star. 
In the context of this large number of analyses often conducted on small samples and limited to a particular subset of targets, the two larger surveys had an~important role. They were conducted with two of the most important high-contrast imagers currently in activity, namely GPI@Gemini, and~SPHERE@VLT. The~Gemini Planet Imager Exoplanets Survey (GPIES \cite{nielsenetal2019aj158_13} has observed 521 nearby and young stars 
further confirming that the frequency of planetary mass companions declines for separations larger than 10 au. They also found that the frequency of planets is dependent on the mass of the host star, with values of 5.3--13.9\% for planets in the mass range between 5 and 13 M$_\text{J}$, around stars with a mass larger than 1.5 M$_\odot$ compared to the lower frequency of 2.1--5.4\% for the same type of objects around lower mass stars. They also concluded that the main formation mechanism for wide-separation planets should be core-accretion, while for the brown dwarf, the main mechanism should be gravitational instability. 
Similar results were obtained by the intermediate sample of 150 stars out of the foreseen 400 targets for the full survey for the SHINE survey performed using SPHERE@VLT and described in \citet{viganetal2021aa651_a72}. They found a frequency of 13.3--36.5\% for planets with a mass between 1 and 75 M$_\text{J}$ at separations between 5 and 300 au from the host star to be compared to the much lower frequency of 3--10.5\% for such objects around FGK stars. On~the contrary, the~frequency for such objects around M stars was of the order of 5.5--25.5\%. The~authors concluded that a planet-like formation pathway dominates for more massive host stars (B and A spectral types), while for the M spectral type, the detection is dominated by brown dwarf binaries. A~combination of these two pathways is instead valid for stars of F, G, and~K spectral types. 
The results of large surveys, summarized in Table\ \ref{tab:occurrencedi}, were confirmed by  smaller surveys come after them. The~Scorpion Planet Survey performed exploiting SPHERE was dedicated to searching planets around 84 A-spectral type stars in the Sco OB2 association. Their results confirmed the larger occurrence of planets at lower separation from the host star and the larger probability of finding planetary mass companions at separations larger than 100 au when observing around A stars. The~survey Imaging Survey for Planets around Young stars (ISPY, \citep{cugnoetal2023aa669_A145}) exploited NACO to observe 45 young stars hosting a known protoplanetary disk. For~these systems, they found a frequency for low-mass companions in line with what they found for more evolved systems.
In very recent times, the possibility that the frequency of exoplanets at large separation from their host star could have been underestimated was suggested by \citet{grattonetal2023natco14_6232}. These authors analyzed 30 stars in the $\beta$-Pic moving group coupling results from direct imaging with those from other techniques like, e.g.,~astrometry, concluding that at least 20 of these targets could host Jupiter-like planets, which suggests that the frequency of such objects could be much larger than expected by previous~analysis.

\begin{table}[H]
\caption{Outcome statistics of some of the surveys listed in Table\ \ref{tab:revhcisurvey2}.}\label{tab:occurrencedi}

\begin{adjustwidth}{-\extralength}{0cm}
\begin{tabularx}{\fulllength}{CCCCCCCC}
\toprule
\textbf{Survey  }           & \textbf{Stars Sp}& \textbf{a}                             & \textbf{M}\boldmath{$_\text{p}$}                   &       \textbf{D}     &  \textbf{Conf. Lev.}   &  \textbf{Frequency}& \textbf{Ref.}     \\
                        &               & \textbf{(au) }                       &  \textbf{(M\boldmath{$_\text{J})$}}                &    \textbf{ (pc)}   &        \textbf{(\%) }      &   \textbf{ (\%)}    &                \\
\midrule

HMPES, KA07& GKM     &  $\geq$30              & $\leq$5 &   $-$ 
     &     $-$            &    <5       &   \citep{kasperetal2007aa472_321} \\        
\midrule
HMPES, BI07& FGKM            &  $22< \text{a} < 507$ & >4 &   $-$     &     95             &    20\ $^\text{a}$      
 &\citep{nielsenandclose2010apj717_878}    \\
                        &                     &  $21< \text{a} < 479$ &                                     &             &                      &               &                                                                   \\
                        &                     &  $82< \text{a} < 276$ &                                    &             &                      &               &                                                                   \\
\midrule
HE10               & FGK      &  $30< \text{a} < 94$ &   >5 &  <11.2 &  90             &  <50     & \citep{heinzeetal2010apj714_1570}      \\
                        &              &  $22< \text{a} < 100$ &   $10$ &               &                  &  <15     &   \\
                        &              &  $8< \text{a} < 100$ &   $20 $ &                 &                  &  <25      &   \\                        
\midrule 
JA11                & BA        & <300                      &  <100                        &   $-$        &   99            & <30     & \citep{jansonetal2011apj736_89}   \\
\midrule
IDPS                &  AF       &   $5< \text{a} < 320$ &   $3<\text{M}_\text{P} < 14$ &       $-$          &      68    &  5.9--18.8           & \citep{viganetal2012aa544_A9}  \\                          
                        &               &   $20< \text{a} < 300$ &   $0.5<\text{M}_\text{P} < 14$ &                &               &  $1.0^{+2.80}_{-0.70}$      & \citep{Galicheretal2016aa594_A63}  \\  
 \midrule
 BI13                &BAFGKM  &  $10< \text{a} < 150$ &  $1<\text{M}_\text{P} < 20$ &     $-$      &       95      &  <18  $^\text{b}$     & \citep{billeretal2013apj777_160}  \\                                            
                        &                  &                                  &                                              &                &       95.4      &  <6  $^\text{c}$      &   \\   
                        &                  &  $10< \text{a} < 50$ &                                               &                &       95.4      &  <21  $^\text{b}$      &   \\    
                        &                  &                                  &                                              &                &       95.4      &  <7  $^\text{c}$      &   \\                                       
\midrule 
RA13               &  AF            &   $1< \text{a} < 1000$ & $1\leq \text{M}_\text{P} \leq 13$ &       $-$      &       68      &  10.8--24.8      &\citep{rameauetal2013aa553_A60}   \\  
\midrule  
SEEDS, GDPS,BI13&BAFGKM & $10< \text{a} < 100$ & $5\leq \text{M}_\text{P} \leq 70$ &  $-$      &       68      &  1--3.1      &\citep{brandtetal2014apj794_159}   \\ 
\midrule
IDPS                        &BAFGKM & $20< \text{a} < 300$ & $0.5\leq \text{M}_\text{P} \leq 14$ &   $-$      &     $-$        &  $1.05^{+2.80}_{-0.70}$ $^{\text{d}}$      &\citep{Galicheretal2016aa594_A63}   \\ 
                                 &                &                                   &                                                       &            &             & $2.30^{+5.95}_{-1.55}$ $^{\text{e}}$       &    \\ 
\midrule
MASSIVE                 & M             &$8< \text{a} < 400$ & $2\leq \text{M}_\text{P} \leq 14$ &     $-$    &          68  &  $2.3^{+2.9}_{-0.70}$      &\citep{lannieretal2016aa596_A83}   \\ 
                                 &                &                                   & $2\leq \text{M}_\text{P} \leq 80$  &            &             & $4.4^{+3.2}_{-1.3}$      &    \\ 
\midrule
LEECH                     & FGK        & $5< \text{a} < 50$ & $7\leq \text{M}_\text{P} \leq 10$ &      $-$  &     $-$  &  $\leq$90     &\citep{stoneetal2018aj156_286}   \\ 
\midrule
WEIRD, PSYM-WIDE& BAFGKM & $5< \text{a} < 5000$ & $1\leq \text{M}_\text{P} \leq 20$ &    $-$    &     $-$    &  $<5.2$ $^{\text{f}}$       &\citep{baronetal2019aj158_187}   \\ 
                                 &                &                                   &                                                       &            &             & $11^{+11}_{-5}$ $^{\text{g}}$      &    \\ 
\midrule
GPIES                      & BAFGKM & $10< \text{a} < 100$ & $5\leq \text{M}_\text{P} \leq 13$ &    $-$     &     $-$    &  $9^{+5}_{-4}$ $^{\text{h}}$      &\citep{nielsenetal2019aj158_13}   \\ 
                                 &                &   $10< \text{a} < 100$ & $13\leq \text{M}_\text{P} \leq 80$ &            &             &  $0.8^{+0.8}_{-0.5}$ $^{\text{h}}$       &    \\ 
\midrule
SPHERE/SHINE      &   BA         & $5< \text{a} < 300$ & $1\leq \text{M}_\text{P} \leq 75$     &   $-$      &    $-$     &  $23^{+13.5}_{-9.7}$      &\citep{viganetal2021aa651_a72}   \\ 
                                 &    FGK      &                                   &                                                    &            &             &  $5.8^{+4.7}_{-2.8} $    &    \\ 
                                 &    M          &                                   &                                                    &            &             &  $12.6^{+12.9}_{-7.1}$      &    \\ 
\bottomrule
\end{tabularx}
\end{adjustwidth}
\noindent{\footnotesize{Notes: a:  The authors give different results for different planetary spectra models: \citet{baraffeetal2003aa402_701, burrowsetal2003ApJ596_587, fortneyetal2008apj683_1104} respectively. b: using DUSTY models \citep{baraffeetal2002aa382_563B}; c: using COND models \citep{baraffeetal2003aa402_701}; d: Considering an uniform distribution; e: Using a power law; f: Using cold start planetary models; g: Using hot start planetary models; h: with host stars mass $<1.5$\ M$_\odot$.}}
\end{table}%

\section{Perspective and~Conclusions}
\label{sec:conc}
In this work, we presented the recent advancement in discovering substellar objects using HCI. Up~to now, the number of detections (a few tens of planets) is quite limited  with respect to the thousands found with indirect techniques like the radial velocity (RV) or the transit technique. With~the current instrumentation, the~HCI method is able to explore large separations from the host star where the expected number of massive companions is \mbox{low \citep{viganetal2021aa651_a72}}, while the peak of the distribution of such objects is expected at separations lower than 10~au \citep{nielsenetal2019aj158_13}. Nevertheless, HCI  is fundamental for finding massive young exoplanets at large separations (>10~au) from the host star, a~niche not covered by other techniques, and it hints at these objects' formation mechanisms. As~a matter of fact, the~fundamental HCI contribution in testing, constraining, and~disentangling the planetary formation mechanisms (core accretion, gravitational instability into a disk) is recognized by most of the authors already cited (e.g., \citep{bowler2016pasp128_j2001, currieetal2023ASPC534_799, viganetal2021aa651_a72}).
Furthermore, this technique allows for the atmospheric characterization of these companions, detecting the photons from them at different wavelengths. In~some cases, it has also been possible to obtain spectra at low or medium resolution of such objects, which allows us to determine the presence of molecules like, e.g.,~H$_2$O and CH$_4$ (e.g., \citep{currieetal2023ASPC534_799}). \par

With the current instrumentation, the~only way to access separations <10 au is to observe targets at low distances (<50~pc) from the Sun; however, just a low number of such objects are as young (a few tens of Myr) as needed for a successful observation with HCI.  Coupling the HCI with one or more different techniques, e.g.,~RV and astrometric measurements, can improve the scientific results of the programs aimed at searching and characterizing planetary objects. Firstly, this can improve the probability of detecting a low-mass companion with respect to the expected probability as opposed to the blind surveys conducted up to now, for which the observations of a sample of hundreds of targets were needed to detect a bunch of substellar objects (both BDs and planets). Indeed, informed surveys are limited to targets for which the probable presence of a companion is indicated by the results of other techniques, which can enhance the success expectation of each single observation. 
Secondly, coupling different techniques can help in constraining the physical characteristics of the companions, including their dynamical masses and their orbital parameters. The~definition of the dynamical mass of the companion, at~the moment possible in a few cases, also allows us to calibrate the atmospheric models that, up~to now, are used for defining the photometric mass of the directly detected~companions. \par

In addition, the~predicted gain, in contrast with the future ELTs' novel instrumentation, can reach, at best, $\sim 10^{-8}$ at an angular separation of about 0.3 arcsecs (see Figure\ \ref{fig:instrcontrast}), allowing the observation of reflected light planets like, for example, Jupiter or planets caught with RV technique. The~possibility of reaching closer angular separations with HCI will allow us to search into the habitable zone of F, G, and early K stars to detect giant planets with possible habitable moons or, in~one more optimistic view, to distinguish technosignatures such as, for~example, the~flux ratio of a faint laser signal to residual scattered light around the star as suggested by \citet{videsetal2019AJ158_207}. 
Much more promising is the suite of future space missions, which will be dedicated to host space coronagraphic observatories like the Roman Telescope, whose performance can reach $\sim 10^{-9}$ at an angular separation of 0.5 arcsecs sufficient to resolve a planet like Jupiter orbiting a star at 10 pc (see Figure\ \ref{fig:instrcontrast}). Since the middle of the last century, NASA and ESA have foreseen space-based large aperture mission concepts that will reach deeper contrasts at closer angular separations, operating at a wavelength from UV to IR like LUVOIR (NASA; \citep{2021BAAS...53d.332R}), HabEx (NASA; \citep{2020arXiv200106683G}), which have been merged in one mission named Habitable Worlds Observatory (HWO; \citep{vaughanetal2023mnras524_5477, starketal2024JATIS_10c4006}), and~LIFE (an interferometric mission) (ESA; \citep{quanzetal2022exa54_1197}) should then allow for the detection and the characterization of exoplanetary systems down to earth mass objects around nearby stars. Their goal will also be to detect possible signs of the presence of life (biosignatures). 

These instruments, together with those of the ELTs, will allow the HCI techniques to make a huge forward leap both in detection and characterization capabilities, which will allow them to equal and, in some cases, outperform the most used indirect~methods.

\vspace{6pt} 



\authorcontributions{Conceptualization, R.C. and~D.M.; methodology, R.C. and D.M.; investigation, R.C. and D.M.; writing---original draft preparation, R.C. and D.M.; writing---review and editing, R.C. and D.M.; supervision, R.C. All authors have read and agreed to the published version of the~manuscript.}

\funding{This research received no external~funding.}

\institutionalreview{Not applicable.}

\informedconsent{Not applicable.}

\dataavailability{No new data were created or analyzed in this study. Data sharing is not applicable to this article.}


\conflictsofinterest{The authors declare no conflicts of~interest.} 


\abbreviations{Abbreviations}{
The following abbreviations are used in this manuscript:\\
\vspace{-12pt}
\noindent 
\begin{longtable}[l]{@{}ll}
4QPM & 4-quadrant Phase-Mask \\
ADI     & Angular Differential Imaging\\
AGPM & Annular Groove Phase Mask\\
ALMA & Atacama Large Millimeter/submillimeter Array\\
ANDROMEDA & ANgular Differential OptiMal Exoplanet Detection Algorithm\\
APLC  &Apodized-Pupil Lyot Coronagraph\\
au      & astronomical unit \\
BLC  & Band-Limited Coronagraph \\
DM   & Deformale Mirror\\
DMS & Deformable Mirror Secondary\\
ELT & Extremely Large Telescope \\
EMCCDs & Electron-Multiplying Charge-Coupled Devices\\
ERIS &Enhanced Resolution Imaging Spectrograph\\
ERSP & Early Release Science Program \\
FMMF & Forward Model Matched Filter \\
FoV & Field of View \\
FRIDA & inFRared Imager and Dissector for Adaptive optics\\
GPI & Gemini Planet Imager \\
GTC & Gran Telescopio Canarias \\
GTCAO & GTC Adaptive Optics system \\
HabEx & Habitable Exoplanets observatory \\
HCI   & High-Contrast Imaging \\
HARMONI & High Angular Resolution Monolithic Optical and Near---Infrared IFS\\
HST & Hubble Space Telescope\\
HWO & Habitable Worlds Observatory\\
IFS  & Integral Field Spectrograph \\
IRDIS &Infra-Red Dual-beam Imager and Spectrograph\\
IWA & Inner Working Angle \\
KL & Karhunen--Loeve \\
KLIP& Karhunen--Loeve Image Projection \\
JWST & James Webb Space Telescope\\
LGS & Laser Guide Stars \\
LOCI & Locally Optimized Combination of Images\\
LTAO & Laser Tomography Adaptive Optics\\
LUVOIR &  Large Ultraviolet Optical Infrared Surveyor \\
MAYONNAISE & Morphological Analysis Yielding separated Objects in Near infrAred usIng \\
                         & Sources Estimation\\
METIS &  Mid Infrared ELT Imager and Spectrograph \\
MICHI & Mid-IR Camera, High-disperser and IFU spectrograph \\
MIRI & Mid-Infrared Instrument  \\
MLOCI & Matched LOCI \\
MUSE & Multi Unit Spectroscopic Explorer \\
NGS & Natural Guide Star\\
NIRCam & Near-infrared Camera  \\
NIRSpec & Near-infrared Spectrograph  \\
NIRISS& Near-Infrared Imager and Slitless Spectrograph  \\
OM & Observing Mode\\
PACO & PAtch COvariance \\
PCA & Principal Component Analysis\\
PCS & Planetary Camera and Spectrograph\\
PMa & Proper Motion Anomaly\\
PMC & Planetary Mass Companion\\
PSI & Planetary System Instrument \\
PSF & Point Spread Function \\
RSM & Regime Switching Model\\
RV & Radial Velocity\\
SAXO & SPHERE extreme AO system\\
SCAO& Single-Conjugate Adaptive Optics\\
SCExAO& Subaru Coronagraphic Extreme Adaptive optics\\
SDI & Spectral Differential Imaging \\
SINFONI & Spectrograph for Integral Field Observation in the Near-Infrared \\
SPHERE & Spectro-Polarimetric High-contrast imager for Exoplanets REsearch  \\
SPIFFI & Spectrometer for Infrared Faint Field Imaging \\
STIM & standardized trajectory intensity mean \\
SVD & Single Value Decomposition \\
TLOCI & Template LOCI \\
TRAP & Temporal Reference Analysis of Planets\\
VIP & Vortex Imaging Processing \\
WFS & WaveFront Sensor\\
ZIMPOL & Zurich Imaging Polarimeter\\
\end{longtable}}



\begin{adjustwidth}{-\extralength}{0cm}
\printendnotes[custom] 

\reftitle{References}

\PublishersNote{}
\end{adjustwidth}
\end{document}